\documentclass[11pt,a4paper]{article}
\usepackage{epsfig}

\newcommand{\unit}[1] {\mbox{\hspace{0.3em}\rm #1}}
\newcommand{\gev}{GeV$^2$}
\newcommand{\asmz}{\alpha_s(M_Z^2)}

\newcommand{\msbar}{\mbox{$\overline{\rm{MS}}$}\ }

\newcommand{\Gl}[1]{Eq.~(\ref{#1})}
\newcommand{\Ab}[1]{Fig.~\ref{#1}}
\newcommand{\Ta}[1]{Table~\ref{#1}}
\newcommand{\Se}[1]{Sec.~\ref{#1}}

\newcommand{\de}{\delta}

\newcommand{\La}{\Lambda}

\def\3{\ss}                                                                                        
                                                                           
\renewcommand{\author}{ }      

\begin{document}

\title {\begin{flushright}{\large DESY--98--121}  \end{flushright}
\vspace{2cm}
\bf\LARGE  ZEUS Results on the Measurement and Phenomenology of
     $F_2$ at Low $x$ and Low $Q^2$ \\
\vspace{1cm}}
                    
\author{ZEUS Collaboration}
\date{}

\maketitle
\begin{abstract}
\noindent
Measurements of the proton structure function $F_2$ for 
$0.6<Q^2<17 \unit{GeV}^2$ and $1.2\times 10^{-5} <x<1.9\times 10^{-3}$
from ZEUS 1995 shifted vertex data are presented. 
From ZEUS $F_2$ data the slopes $dF_2/d\ln Q^2$ at fixed $x$ 
and $d\ln F_2/d\ln(1/x)$ for $x<0.01$ at fixed $Q^2$ are derived. 
For the latter E665 data are also used. The transition region in $Q^2$
is explored using the simplest non-perturbative models and NLO QCD. 
The data at very low $Q^2$ $\leq 0.65\unit{GeV}^2$ are described 
successfully by a combination of generalised vector meson dominance and 
Regge theory. From a NLO QCD fit to ZEUS data the gluon density in the 
proton is extracted in the range $3\times 10^{-5}<x<0.7$.
Data from NMC and BCDMS constrain the fit at large $x$. 
Assuming the NLO QCD description to be valid down to $Q^2\sim 1\unit{GeV}^2$, 
it is found that the $q\bar{q}$ sea distribution is still 
rising at small $x$ and the lowest $Q^2$ values whereas the gluon 
distribution is strongly suppressed.

\end{abstract}

\pagestyle{plain}    
\thispagestyle{empty}
\clearpage

%
%
%
%
\pagenumbering{Roman}                                                                              
                                                   %
\begin{center}                                                                                     
{                      \Large  The ZEUS Collaboration              }                               
\end{center}                                                                                       
  J.~Breitweg,                                                                                     
  S.~Chekanov,                                                                                     
  M.~Derrick,                                                                                      
  D.~Krakauer,                                                                                     
  S.~Magill,                                                                                       
  D.~Mikunas,                                                                                      
  B.~Musgrave,                                                                                     
  J.~Repond,                                                                                       
  R.~Stanek,                                                                                       
  R.L.~Talaga,                                                                                     
  R.~Yoshida,                                                                                      
  H.~Zhang  \\                                                                                     
 {\it Argonne National Laboratory, Argonne, IL, USA}~$^{p}$
% ...
\newpage                                                                                           
% ...
$^{  17}$ now at ATLAS Collaboration, Univ. of Munich \\                                           
% ...
$^{  30}$ partially supported by the Foundation for German-Russian Collaboration                   
DFG-RFBR \\ \hspace*{3.5mm} (grant no. 436 RUS 113/248/3 and no. 436 RUS 113/248/2)\\              
% ...
                                                           %
                                                           %
\newpage   
                                                           %
                                                           %
\begin{tabular}[h]{rp{14cm}}                                                                       
$^{a}$ &  supported by the Natural Sciences and Engineering Research                               
          Council of Canada (NSERC)  \\                                                            
$^{b}$ &  supported by the FCAR of Qu\'ebec, Canada  \\                                            
$^{c}$ &  supported by the German Federal Ministry for Education and                               
          Science, Research and Technology (BMBF), under contract                                  
          numbers 057BN19P, 057FR19P, 057HH19P, 057HH29P \\                                        
$^{d}$ &  supported by the MINERVA Gesellschaft f\"ur Forschung GmbH,                              
          the German Israeli Foundation, the U.S.-Israel Binational                                
          Science Foundation, and by the Israel Ministry of Science \\                             
$^{e}$ &  supported by the German-Israeli Foundation, the Israel Science                           
          Foundation, the U.S.-Israel Binational Science Foundation, and by                        
          the Israel Ministry of Science \\                                                        
$^{f}$ &  supported by the Italian National Institute for Nuclear Physics                          
          (INFN) \\                                                                                
$^{g}$ &  supported by the Japanese Ministry of Education, Science and                             
          Culture (the Monbusho) and its grants for Scientific Research \\                         
$^{h}$ &  supported by the Korean Ministry of Education and Korea Science                          
          and Engineering Foundation  \\                                                           
$^{i}$ &  supported by the Netherlands Foundation for Research on                                  
          Matter (FOM) \\                                                                          
$^{j}$ &  supported by the Polish State Committee for Scientific                                   
          Research, grant No.~115/E-343/SPUB/P03/002/97, 2P03B10512,                               
          2P03B10612, 2P03B14212, 2P03B10412 \\                                                    
$^{k}$ &  supported by the Polish State Committee for Scientific                                   
          Research (grant No. 2P03B08614) and Foundation for                                       
          Polish-German Collaboration  \\                                                          
$^{l}$ &  partially supported by the German Federal Ministry for                                   
          Education and Science, Research and Technology (BMBF)  \\                                
$^{m}$ &  supported by the Fund for Fundamental Research of Russian Ministry                       
          for Science and Edu\-cation and by the German Federal Ministry for                       
          Education and Science, Research and Technology (BMBF) \\                                 
$^{n}$ &  supported by the Spanish Ministry of Education                                           
          and Science through funds provided by CICYT \\                                           
$^{o}$ &  supported by the Particle Physics and                                                    
          Astronomy Research Council \\                                                            
$^{p}$ &  supported by the US Department of Energy \\                                              
$^{q}$ &  supported by the US National Science Foundation \\                                       
\end{tabular}                                                                                      
                                                           %
                                                           %
\newpage
\setcounter{page}{1}
\pagenumbering{arabic}

\section{Introduction}
\label{sec:intro}

The rapid rise of the proton structure function
$F_2$ at low $x$, measured at HERA \cite{heraf2}, continues to generate 
a lot of interest. In particular the persistence of the strong rise 
to small values of $Q^2$ and the apparent success of the  perturbative QCD 
(pQCD) description of the data down to $Q^2$
values approaching $1\unit{GeV}^2$ raise new challenges for our 
understanding of QCD. HERA also allows study of the `transition region' as
$Q^2\to 0$ in which pQCD must break down. The theoretical context
for our study of pQCD and the transition region is outlined in 
\Se{sec:phenom}.  

With data taken during the 1995 HERA run the ZEUS experiment has 
achieved a significant increase in the kinematic coverage for low $x$ 
and low $Q^2$ inelastic neutral current positron-proton scattering. 
The coverage for $Q^2$ between 0.11 and $0.65 \unit{GeV}^2$ was made
possible with the installation of a small electromagnetic sampling
calorimeter, the Beam Pipe Calorimeter (BPC), at small positron
scattering angles and results on the proton structure
function $F_2$ and the total $\gamma^*p$ cross-section 
have been published \cite{ZBPC}.
In \Se{sec:svxdat} of this paper we report on further measurements of $F_2$ in
the $Q^2$ range between 0.6 and $17 \unit{GeV}^2$. The data were
obtained from runs in which the interaction point was
shifted away from the main rear calorimeter thus extending
its small-angle coverage for scattered positrons.  These data fill
the gap in $Q^2$ between the BPC and the 1994 ZEUS 
measurements \cite{ZSVX94,ZF2}. Taking all three data sets together,
the ZEUS experiment has measured $F_2$ over the kinematic region
$0.1<Q^2<5000 \unit{GeV}^2$, $2 \times 10^{-6}<x<0.5$. The
coverage of the kinematic plane by the ZEUS data sets is shown
in \Ab{fig:kinreg}. 

The very low $Q^2$ data are discussed in \Se{sec:lowq2reg} using generalised 
vector dominance and Regge theory and it
is established that the ZEUS data with $Q^2\leq0.65\unit{GeV}^2$ are
well described by such approaches. In \Se{sec:results} the 
slopes $dF_2/d\ln Q^2$ at fixed $x$ and $d\ln F_2/d\ln(1/x)$ at 
fixed $Q^2$ are derived from the combined ZEUS data sets. In \Se{sec:qcd}
the ZEUS $F_2$ data, together with fixed target data at large $x$, 
are fit using next to leading order (NLO) QCD to determine the gluon 
momentum density. The 
increased range and precision of the ZEUS $F_2$ data allow a more precise
extraction of the gluon density at low $x$ compared to 
our earlier results using the 1993 ZEUS data \cite{zg93}.
In \Se{sec:loq2qcd} the properties of the pQCD description of the $F_2$ 
and slopes data at low $Q^2$ are explored in more detail. 
Our conclusions are summarised in \Se{sec:concl}. Tables containing
$F_2$ values and other data are given in the Appendix.

\section{Phenomenology of the low $Q^2$ region}
\label{sec:phenom}

We use NLO pQCD and the simplest non-perturbative models to 
explore the transition region in $Q^2$.
The standard NLO DGLAP equations \cite{dglap} give the $Q^2$ evolution of 
parton densities, but do not prescribe their functional form in 
$x$~\footnote{Bjorken $x$ and the negative squared momentum transfer, $Q^2$, 
are defined in \Se{sec:kinrec}.} at
the starting scale $Q^2_0$. At $Q^2_0$, the small $x$ behaviour of
parton momentum densities $xf(x)$ may be characterised by the 
exponent $\delta$ where $xf(x)\approx Ax^{\delta}$. For
$\delta \geq 0$ a parton momentum density either tends to zero or is 
constant as $x\to 0$ (non-singular) while for 
$\delta < 0$ a parton density increases as $x\to 0$ (singular).
One way to understand the rise of $F_2$
at low $x$ is advocated by Gl\"uck, Reya and Vogt (GRV) \cite{grv,grv94}
who argue that the starting scale for the evolution of the parton
densities should be very low ($0.34$ GeV$^2$) and at the
starting scale the parton density functions should be non-singular. 
For $Q^2 > 1 \unit{GeV}^2$, the observed rise in $F_2$
is then generated dynamically through the DGLAP evolution equations.
On the other hand, at low $x$ one might expect that the DGLAP
equations break down because of large $\ln(1/x)$
terms that are not included. Such terms are taken into account by
the BFKL formalism \cite{bfkl}, which in leading order predicts a
rising $F_2$ at low $x$. The rise comes from a singular gluon density
with $\delta$ in the range $-0.3$ to $-0.5$. Recent work on BFKL at 
NLO has shown that the corrections to the LO value for $\delta$ are 
large \cite{nlobfkl} and reduce the predicted rise in 
$F_2$, though quite how large the reduction should be is still under discussion
\cite{rossd}. Clearly accurate experimental results on 
$F_2$ and $\delta$ at low $x$ are of great interest.
More details on the many alternative pQCD approaches to the low $x$ region
may be found in \cite{dis96-98,cddrev}.

As $Q^2$ decreases $\alpha_S$ increases and pQCD will eventually
break down. Then non-perturbative models must be used to describe the data.  
At low $x$ the lifetime of the virtual photon in the proton rest frame
is large compared to the $\gamma^*p$ interaction time \cite{Ioffe}.  
Inelastic $e^\pm p$  scattering may then be viewed as $\gamma^*p$ 
scattering, with the total $\gamma^*p$ cross-section given by~\footnote{
Considering virtual photon exchange only. Since we are working at small 
$x$, terms depending on $M_N^2x^2$ have been ignored.} 
\begin{equation}
\sigma_{\rm tot}^{\gamma^*p}(W^2,Q^2)\equiv
        \sigma_T+\sigma_L \approx \frac{4\pi^2\alpha}{Q^2} F_2(x,Q^2),
\label{eqn:f2sig}
\end{equation}
where $W \approx \sqrt{Q^2/x}$ is the centre-of-mass energy of the 
$\gamma^*p$ system and $\sigma_T$ and $\sigma_L$ are the cross-sections
for transversely and longitudinally polarised virtual photons respectively.  
We consider two non-perturbative approaches, the vector meson dominance 
model (VMD) and Regge theory. 

VMD relates the hadronic interactions of the 
photon to a sum over interactions of the $\rho^0,\omega$ and $\phi$ vector 
meson states \cite{sak60,bauer}. To accommodate
deep inelastic scattering data the sum has to be extended to an infinite 
number of vector mesons giving the generalised
vector dominance model (GVMD) \cite{sak72}. Following the assumptions
in \cite{sak72}, $\sigma_T(W^2,Q^2)$ is related to $\sigma^{\gamma p}(W^2)$,
the total photoproduction cross-section by
\begin{equation}
\sigma_T(W^2,Q^2)=
 \left[{r_CM_0^2\over (M_0^2+Q^2)}+\sum_{V=\rho^0,\omega,\phi}
 {r_VM_V^4\over (M_V^2+Q^2)^2}\right]\sigma^{\gamma p}(W^2),
\label{eqn:gvmd}
\end{equation}
where $M_0$ is the lower cutoff of the continuum vector states and
$r_C,r_V$ are constants satisfying the normalisation condition 
$r_C+\sum_V r_V=1$ at $Q^2=0$. A similar expression to \Gl{eqn:gvmd} may be 
written for $\sigma_L(W^2,Q^2)$,
but with additional $Q^2$ dependence to ensure that it vanishes as
$Q^2\to 0$ \cite{sak72}. The GVMD approach has recently been
revived in the context of low $Q^2$ HERA data by Schildknecht and
Spiesberger \cite{schsp}. We use a simplified form of GVMD to study the 
consistency of the data in the ZEUS BPC region and its extrapolation to 
$Q^2=0$.

Regge theory \cite{regge,levin} provides a framework in which 
the energy dependence of hadronic total cross-sections is of the
form $\sigma\sim \sum_r\beta_r s^{\alpha_r-1}$ where 
$\sqrt{s}$ is the centre-of-mass energy, $\alpha_r$
the intercept of the Regge trajectory and $\beta_r$ a process
dependent constant. The $\alpha_r$ are universal and can in principle 
be determined from the spectrum of meson states. However, for 
the dominant trajectory describing total cross-sections at high energies 
(known as the Pomeron), this has not yet been possible. The Pomeron 
intercept, $\alpha_P$, is determined by fitting high energy total 
cross-section data. Donnachie and Landshoff (DL) \cite{dlone} have used a 
two component Pomeron+Reggeon approach to give a good description
of hadron-hadron and photoproduction total cross-section data 
over a wide range of energies with $\alpha_P$ of about 1.08. 
They have extended their approach to $\gamma^*p$ total cross-sections 
\cite{dltwo} by keeping the Regge intercept independent of $Q^2$ but assuming
a simple $Q^2$ dependence for the coupling term, which
becomes $\beta_r m_r^2/(m_r^2+Q^2)$ where $m^2_r$ is
again determined by fitting to data. 

Neither the non-perturbative VMD and DL approaches nor pQCD can be expected
to describe the $Q^2$ behaviour of $F_2$ over the complete range from 
photoproduction to very large $Q^2$ deep inelastic scattering (DIS).
Many models combining various aspects of these 
approaches have been applied to the data in the transition
region \cite{cddrev,bkrev,levy}. A comparison of the ZEUS BPC data with 
some of the models was given in \cite{ZBPC}. We use the DL Regge model 
\cite{dltwo} and the 1994 parton densities (GRV94) of the 
GRV group \cite{grv94} as `benchmarks' in this paper as
they were available before the recent precision measurements were made.

\section{Measurement of $F_2$ with shifted vertex data}
\label{sec:svxdat}

The shifted vertex data correspond to an integrated luminosity of 
$236 \unit{nb}^{-1}$ taken in a special running period in which 
the nominal interaction point was offset in the proton beam 
direction by $+70\unit{cm}$,~\footnote{The ZEUS coordinate system is defined 
as right handed with the $Z$ axis pointing in the proton beam direction, 
and the $X$ axis horizontal, pointing towards the centre of HERA. The 
origin is at the unshifted interaction point.} i.e.
away from the main rear calorimeter (RCAL).
The measurement follows previous analyses described in more detail in
\cite{ZSVX94, ZF2}. However, compared to the earlier shifted
vertex analysis \cite{ZSVX94}, for the 1995 data taking period the
RCAL modules above and below the beam were moved closer to the beam, thus
extending the shifted vertex $Q^2$ range down to $0.6 \unit{GeV}^2$.
The basic detector components used are the
compensating uranium calorimeter (CAL), which has an energy resolution, 
as measured in the test beam, of $\sigma/E=18\%/\sqrt{E({\rm GeV})}$ for 
electromagnetic particles and $\sigma/E=35\%/\sqrt{E({\rm GeV})}$ for 
hadronic particles. The tracking chamber system is used to determine
the position of the event vertex.  The small angle rear tracking 
detector (SRTD) consists of horizontal and vertical scintillator strips 
covering the region around the RCAL
beam hole, i.e. the region of positron scattering angles of low $Q^2$ events.
It is also used as a preshower detector for the RCAL and has a position 
resolution of $0.3 \unit{cm}$. 
The luminosity is determined from the positron-proton bremsstrahlung
$ep\rightarrow ep\gamma$ where the radiated photon is measured in a
lead-scintillator calorimeter (LUMI) positioned at $Z=-107\unit{m}$.
There is an associated electron calorimeter (LUMI-E), positioned at
$Z=-35\unit{m}$, which is used for tagging photoproduction events.
The uncertainty in the luminosity measurement is $1\%$. 

\subsection{Monte Carlo simulation}
\label{sec:svxmc}

Monte Carlo (MC) events are used to correct for the detector acceptance,
resolution and the effect of initial state radiation. In the framework
of DJANGO \cite{django} the generator HERACLES \cite{heracles} is used
to simulate neutral current DIS events including first order
electroweak radiative effects. The hadronic final state is simulated using the
ARIADNE \cite{ariadne} program which implements the colour-dipole model.
A parameterisation of the $F_2 $ structure functions 
based on the results published in \cite{ZF2} is used and $F_L$ is 
set to zero. The MC event sample is 1.6 times that of
the data and the events are passed through the same offline reconstruction 
software as the data. 
Simulated photoproduction background events are 
generated using the program PYTHIA \cite{pythia} with a cross-section given 
by the ALLM \cite{allm} parameterisation.

\subsection{Kinematic reconstruction}
\label{sec:kinrec}
The reaction 
$e^+(k) + p(P) \rightarrow e^+(k') + X$ at fixed squared centre-of-mass 
energy, $s=(k+P)^2$, is described in terms of $Q^2\equiv -q^2= -(k-k')^2$ 
and Bjorken $x = Q^2/(2P\cdot q)$. At HERA $s\approx 4 E_e E_p$, where 
$E_e=27.5\unit{GeV}$ and $E_p=820\unit{GeV}$ denote the 
positron and proton beam energies. The fractional energy transferred
to the proton in its rest frame is $y = Q^2/(sx)$. 

The kinematic variables are reconstructed from the measured energy, $E_e'$,
and scattering angle, $\theta_e$, of the positron
 (the `electron method'),
\begin{displaymath}
Q^2 = 2E_eE_e^\prime\left(1+\cos \theta_e\right)
\,\,\,{\rm and}\,\,\,
y = 1-\frac{E_e^\prime}{2E_e} \left( 1-\cos\theta_e \right).
\end{displaymath}
This method gives the best resolution in the region of interest  
at high $y$ and low $Q^2$.

Scattered positrons are identified by a neural network based algorithm 
\cite{ZF2}, with an efficiency of about 90\% at positron energies of  
$10 \unit{GeV}$, increasing to $100\%$ at $20 \unit{GeV}$.
The measured energy is corrected for energy loss in inactive material in 
front of the CAL using the signals in the SRTD scintillators. 
The uncertainty in the measured energy is estimated to be $2\%$ at
$10\unit{GeV}$ decreasing to $1\%$ at $27.5\unit{GeV}$. 
The positron impact position on the CAL measured with the SRTD 
together with the event vertex position measured with tracks in the CTD gives 
the positron scattering angle $\theta_e$. 
For events outside the fiducial volume of the SRTD, the CAL position 
determination is used. 
For events in which the event vertex cannot be reconstructed,
the vertex is set to the mean value of the vertex distribution.

The variable $y_{\rm JB} = \sum_i E_i (1-\cos \theta_i) /(2 E_e)$, where 
the sum runs over all CAL cells except those belonging to the 
scattered positron, gives a measurement of $y$ with a good resolution at 
low $y$. A cut $y_{\rm JB}>0.04$ is imposed to limit event migrations from 
low $y$, where the resolution of the electron method is poor, into the bins at 
higher $y$.

\subsection{Event selection}
\label{sec:svxcuts}

Data are selected online by a three level trigger system. 
At the first level a certain energy deposit in the CAL is required  
and cuts on the arrival times of particles measured in the
SRTD are imposed. At the second level the condition 
$\delta \equiv \sum_i E_i(1-\cos\theta_i) > 29\unit{GeV} - 2 E_{\gamma}$ 
has to be fulfilled, where 
the sum goes over all CAL cells with energies $E_i$ and polar angles 
$\theta_i$, and $E_{\gamma}$ is the energy   
measured in the LUMI detector. This cut significantly reduces the  
photoproduction background as $\delta=2E_e$ ($55 \unit{GeV}$) 
for a fully contained DIS event. At the third level a full event 
reconstruction is performed. A reconstructed positron with an energy 
greater than $4 \unit{GeV}$ and a CAL impact point outside a 
box of $ 24 (X) \unit{cm} \times 12(Y) \unit{cm}$ centered on the 
RCAL beam hole is 
required. Also, $\delta$ has to be greater than $30 \unit{GeV}$ and 
event times measured in the CAL are required to be consistent with an 
$e^+p$ interaction at the nominal shifted interaction point. 

The offline event selection cuts are: 
\begin{itemize}
\item 
Positron finding as described above, including the requirements on the impact 
point and on $y_{\rm JB} > 0.04$, with a corrected positron energy 
$E_e^\prime > 10 \unit{GeV}$. 
This ensures a high efficiency for positron finding and removes events at very 
high $y$, which suffer from large photoproduction backgrounds.
\item
The positron impact point on the CAL is required to be outside 
a box of $26 \unit{cm} \times 16 \unit{cm}$ around the RCAL beam pipe hole 
to ensure full shower containment in the CAL. 
\item 
$35 \unit{GeV} < \delta < 65 \unit{GeV}$, in order to further reduce 
photoproduction and beam-gas related backgrounds. 
This cut also removes events with hard initial state radiation.
\item 
For events with a tracking vertex, the reconstructed $Z$
coordinate of the vertex is required to lie within 
$40 \unit{cm} < Z_{\rm vertex} < 160 \unit{cm}$. 
The acceptance is extended to large $Z$ values
to accommodate events from satellite bunches, 
i.e. proton bunches that are shifted by
$4.8\unit{ns}$ with respect to the primary bunch crossing time,
resulting in a fraction of $ep$ interactions occurring displaced by
an additional $72\unit{cm}$.
\end{itemize}
A total of 62000 events pass the cuts.

\subsection{Background estimation}

The background from beam-gas interactions is about 1\% as
determined from unpaired positron and proton bunches. 

The main background comes from photoproduction events, where  
the positron escapes along the beam line and a mis-identified 
positron (mainly electromagnetic showers from 
$\pi^0$ decays) is reconstructed in the CAL. 
The amount of this background is determined using the MC simulated 
photoproduction event sample. In total it is a small effect which
is only significant at small values of $E^\prime_e$ and $Q^2$ as shown in 
plots (d) and (e) of \Ab{fig:svxplots}.
In a small fraction of real photoproduction events the 
positron is detected inside the limited acceptance of the LUMI-E 
electron tagger. These events are used to cross check the MC background
estimate. Both results agree within 20\%. 

\subsection{Determination of $F_2$}

In the $Q^2$ range of this analysis the double differential cross-section for 
single virtual-photon exchange in DIS is given by
\begin{equation}
  \frac{d^2\sigma}{dxdQ^2} = 
\frac{2\pi\alpha^2}{x Q^4}
\left[ 
2\left( 1-y \right) + \frac{y^2}{1+R}
\right] 
F_2(x,Q^2) 
\left[ 1 + \delta_r(x,Q^2) \right],
\label{eqn:d2dxdq2}
\end{equation}
where $R$ is related to the longitudinal structure function $F_L$ by 
$R=F_L/(F_2-F_L)$ and $\delta_r$ gives the radiative correction to the 
Born cross-section. For the kinematic range of this analysis $\delta_r$
is at most $10\%$. For $R$ we take values given by the BKS model \cite{BKS}.
An iterative procedure is used to extract the structure function $F_2$.
Data and MC events are binned in the variables $y$ and $Q^2$. 
In a bin-by-bin unfolding procedure the MC differential cross-section is  
adjusted to describe the data using a smooth function for $F_2$. 
The re-weighted MC events are then used to unfold $F_2$ again, 
until after 3 iterations the changes to $F_2$ are below $0.5\%$.
The statistical errors of the $F_2$ values are calculated from the
number of events measured in a bin and the statistical error on the
acceptance calculation from the MC simulation.

The quality of the description of the data by the re-weighted MC is
shown in \Ab{fig:svxplots}, which displays distributions of the 
following quantities: (a) $\delta$ or $E-P_Z$ as defined in \Se{sec:svxcuts};
(b) the $Z$-position of the primary vertex; (c) the positron scattering
angle $\theta_e$; (d) the energy, $E^\prime_e$, of the scattered positron;
(e) $\log_{10}Q^2$; (f) $\log_{10}y$. The agreement between data (filled
circles) and simulated data (open histograms, normalised to the luminosity
of the data) is generally good. 

\subsection{Systematic uncertainties}

The systematic uncertainties of the measured $F_2$ values are determined by 
changing the selection cuts or the analysis procedure in turn and repeating
the extraction of $F_2$. Positive and negative 
differences, $\Delta F_2$, are added in quadrature separately to 
obtain the total positive and negative systematic errors. The systematic 
checks and errors are: 
\begin{itemize}
\item  
A shift of the horizontal and vertical position of the 
 SRTD by $\pm 0.5\unit{mm}$ results in $\Delta F_2/F_2$ of at 
most $\approx 2\%$.
\item 
The two halves of the SRTD are moved with respect to each other 
in the horizontal and vertical directions by $1.0\unit{mm}$, which results 
in $\Delta F_2/F_2$ of $\approx 4\%$.
\item 
The uncertainty in the positron energy calibration is estimated to be
$2\%$ at $10\unit{GeV}$ decreasing to $1 \%$ at $27.5\unit{GeV}$ 
giving a maximum $\Delta F_2/F_2$ of $8\%$.
\item 
The hadron energy scale is uncertain to $\pm 3\%$, causing 
a maximum $\Delta F_2/F_2$ of 2\%.
\item 
The uncertainty in the positron finder efficiency is estimated to be
$2.5\%$ at $10\unit{GeV}$ decreasing to $1 \%$ at $27.5\unit{GeV}$, 
which gives $\Delta F_2/F_2$ of at most 2\%.
\item 
The number of MC events in satellite bunches
is increased by 100\%  and decreased to 50\%.
This leads to a $\Delta F_2/F_2$ less than 2\%.
\item 
The uncertainty in the determination of the photoproduction background  
by MC is estimated to be $30\%$. The resulting maximum $\Delta F_2/F_2$ is 
about 7\% in the highest $y$ bins.
\item 
The vertex finding efficiency is between 75\% and 95\% depending on
the kinematic region. To study the effect of differences in the efficiency
between MC and data all vertices are assigned to the nominal 
interaction point. $\Delta F_2/F_2$ is at most 6\%.
\item
A variation in the box cut  from $26\unit{cm}\times 16\unit{cm}$ 
to $25.6\unit{cm}\times 15.6\unit{cm}$ or $26.4\unit{cm}\times 8.4\unit{cm}$, 
giving a maximum change $\Delta F_2/F_2$ of 7\%.
\end{itemize}

The acceptance for DIS events with a rapidity gap at low $Q^2$ 
is somewhat different from that of non-diffractive events
due to the different energy flow. To
check the effect on $F_2$, the acceptance for diffractive events is
first calculated using a separate diffractive MC.\footnote{A modification
of the ARIADNE MC adjusted to generate rapidity gap events as described
in \cite{ZF2}.}
 Using this and the
measured fraction of rapidity gap events in each bin, the 
acceptance function is recalculated. The largest change to $F_2$ is $2.5\%$. 
The data are corrected for this effect. Half the 
correction value is taken as the estimate of the systematic error, 
reflecting mainly the uncertainty in the fraction of diffractive events.

In addition there is an overall normalisation uncertainty of 1.5\%,
due to the 1\% error in the luminosity measurement and a 1\% uncertainty
in the trigger efficiency, which is not included in the point to point 
systematic error estimate.
 
The $F_2$ data cover the $x$ range $1.2\times 10^{-5} - 1.9\times 10^{-3}$ 
in 12 
bins of $Q^2$ between 0.6 and $17\unit{GeV}^2$ (ZEUS SVX95).
The values for $F_2$ and their systematic errors are given in 
\Ta{tab:svx95} of the Appendix.
\Ab{fig:svtx1} shows the results for $F_2$ as a function of $x$ in the
12 $Q^2$ bins. In the lowest $Q^2$ bin data from ZEUS $F_2$ measurements 
at very low $Q^2$ using the BPC (BPC95) \cite{ZBPC} are shown and at 
larger $Q^2$ those from the ZEUS94 measurements \cite{ZF2}. Also shown
are data from the shifted vertex measurements by H1 (H1 SVX95) \cite{H1_svx95}
and fixed target data from E665 \cite{E665F2}. There is good agreement
between the different ZEUS data sets and between ZEUS, E665 and H1 data
in the regions of overlap. We note that the steep increase of $F_2$ at low 
$x$ observed in the higher $Q^2$ bins softens at the lowest $Q^2$ values. 
The curves shown will be discussed later.

\section{The low $Q^2$ region}
\label{sec:lowq2reg}

We first give an overview of the $Q^2\leq 4.55\unit{GeV}^2$ region.
\Ab{fig:svtx2} shows the ZEUS cross-section data versus $W^2$ 
derived from the SVX95, BPC95 and ZEUS94 $F_2$ values using \Gl{eqn:f2sig}.
Also shown are data from H1 SVX95 and measurements of
the total cross-section for scattering of real photons on protons 
at fixed target \cite{phptot} and HERA energies \cite{phptotHERA}.
The two curves shown are the predictions of the DL Regge model \cite{dltwo} 
and $\sigma_{\rm tot}^{\gamma^*p}$ calculated from the NLO QCD parton
distributions of GRV94 \cite{grv94}. The DL model predicts that the 
cross-section rises slowly with energy $\propto W^{2\lambda}$, 
$\lambda = 1-\alpha_P\approx 0.08$ and this behaviour
seems to be followed by the data at very low $Q^2$ values, although the
normalisation of the DL model is low compared to the ZEUS BPC95
data. Above $Q^2 = 0.65\unit{GeV}^2$, the DL model predicts 
a shallower rise of the cross-section than the data exhibit. 
For $Q^2$ values around $1\unit{GeV}^2$ and above, the GRV94 curves 
describe the qualitative behaviour of the data, namely the  increasing
rise of $\sigma_{\rm tot}^{\gamma^*p}$ with $W^2$, as $Q^2$ increases.
This suggests that the pQCD calculations can account for 
a significant fraction of the cross-section at the larger $Q^2$ values. 

For the remainder of this section we concentrate on non-perturbative
descriptions of the ZEUS BPC95 data ($0.11 < Q^2 < 0.65 \unit{GeV}^2$).
Since the BPC data are binned in $Q^2$ and $y$ we first rewrite the 
double differential cross-section of \Gl{eqn:d2dxdq2}
(dropping the radiative correction factor) as
\begin{equation}
{d^2\sigma\over dydQ^2}=\Gamma\cdot(\sigma_T+\epsilon\sigma_L)
\label{eqn:sigtel}
\end{equation}
where $\displaystyle \sigma_L={Q^2\over 4\pi^2\alpha}F_L$ and
$\sigma_T+\sigma_L$ has been defined by \Gl{eqn:f2sig}. The virtual 
photons have flux $\Gamma=\alpha(1+(1-y)^2)/(2\pi Q^2y)$ and polarisation 
$\epsilon = 2(1-y)/(1+(1-y)^2)$. For the BPC data $\epsilon$ lies in the 
range $0.55-0.99$ but as $s$ is fixed at HERA, $\epsilon$ cannot be varied 
independently of $x$ and $Q^2$. Thus the experimently determined quantity
is the
combination $\sigma_T+\epsilon\sigma_L$. For simplicity we keep only the 
continuum term in the GVMD expression of \Gl{eqn:gvmd}. At a 
fixed $W$ the longitudinal and transverse $\gamma^*p$ cross-sections
are then related to the corresponding cross-section $\sigma_{0}^{\gamma p}$ 
at $Q^2=0$ by
\begin{eqnarray}
\sigma_L(W^2,Q^2) & = & \xi \left[
\frac{M_0^2}{Q^2}\ln\frac{M_0^2+Q^2}{M_0^2} - \frac{M_0^2}{M_0^2+Q^2}
\right]\sigma_{0}^{\gamma p}(W^2)\nonumber\\
\sigma_T(W^2,Q^2) & = & 
\frac{M_0^2}{M_0^2+Q^2}\sigma_{0}^{\gamma p}(W^2),
\label{eqn:vmd}
\end{eqnarray}
where the parameter $\xi$ is the ratio $\sigma_L^{Vp}/\sigma_T^{Vp}$ for
vector meson (V) proton scattering and $M_0$ is the 
effective vector meson mass. Neither $\xi$ nor $M_0$ are given by the model 
and they are usually determined from data. We set $\sigma_L$ to zero
because $\xi$ is expected to be less than one~\footnote{From studies of 
diffractive vector meson production and data on $R(=\sigma_L/\sigma_T)$ 
in inelastic
$ep$ scattering $\xi$ is in the range $0.2 - 0.4$ \cite{bauer}. More recently
ZEUS \cite{zeusvm} has measured $ep\to ep\rho$ at low $Q^2$ and found that
$R$ is about $0.4$ at $Q^2= 0.5\unit{GeV}^2$.} and the factor in the 
square bracket in the expression for $\sigma_L$ is small (for 
$Q^2\leq 0.65\unit{GeV}^2$ and $M_0\approx m_\rho$, it is less than 0.2).  
The $Q^2$ dependence of the BPC data, in 8 bins of $W$ between 104 and 
$251\unit{GeV}$, is fit with a single mass parameter $M_0^2$. The
cross-sections, $\sigma_{0}^{\gamma p}(W^2)$, are also fit at each $W$ giving
a total of 9 parameters. The fit is reasonable
($\chi^2/ndf=38.7/(34-9)=1.55$, statistical errors only) as shown in
the upper plot of \Ab{fig:bpcvmd}. To estimate the systematic errors, the 
fit is first repeated for each systematic check on the BPC data, using data 
and statistical errors corresponding to that change. The systematic errors 
are then determined by adding in quadrature the changes from the nominal 
values of the parameters. As a final check on the stability of the
results, the fit is repeated including the longitudinal term with $\xi=0.4$. 
The resulting changes in the values for the cross-sections are 
less than their statistical errors (more details are given in 
\cite{surrow}). The value obtained for $M_0^2$ is 
$0.53\pm0.04({\rm stat})\pm0.09({\rm sys})\unit{GeV}^2$. The 
resulting extrapolated values of 
$\sigma_{0}^{\gamma p}$ are given in \Ta{tab:sigt} of the Appendix and
shown as a function of $W^2$ in the lower plot of 
\Ab{fig:bpcvmd}, along with measurements from HERA and lower energy
experiments. The BPC $\sigma_{0}^{\gamma p}$ values lie somewhat above 
the direct measurements from HERA. They are also above the prediction of
Donnachie and Landshoff. It should be clearly understood that the 
$\sigma_{0}^{\gamma p}$ values derived from the BPC are not a measurement
of the total photoproduction cross-section but the result of a 
phenomenologically motivated extrapolation. 

The simple GVMD approach just described gives a concise account of the $Q^2$
dependence of the BPC data. To describe the energy dependence of the data 
we use a two component Regge model
\begin{displaymath}
\sigma_{\rm tot}^{\gamma p}(W^2) = A_R (W^2)^{\alpha_R-1}+ 
A_P(W^2)^{\alpha_P-1}
\end{displaymath}
where $P$ and $R$ denote the Pomeron and Reggeon contributions. The
Reggeon intercept $\alpha_R$ is fixed to the value $0.5$ which is 
compatible with the original DL value \cite{dlone}, and recent
estimates \cite{dlpdg,cudell}. Fitting only the Pomeron term to the 
extrapolated BPC data ($\sigma_{0}^{\gamma p}$) gives
$\alpha_P=1.141\pm0.020$(stat)$\pm0.044$(sys). Fitting 
both terms to the real photoproduction data (with $W^2>3\unit{GeV}^2$) 
and BPC $\sigma_{0}^{\gamma p}$ data yields 
$\alpha_P=1.101\pm 0.002$(stat)$\pm0.012$(sys). Including, in addition, 
the two direct measurements from HERA~\cite{phptotHERA} gives 
$\alpha_P=1.100\pm0.002$(stat)$\pm0.012$(sys).
At HERA energies the contribution of the Reggeon term is negligible.
The values of $\alpha_P$ are compatible with the DL value
of 1.08 and the recent best estimate of $1.0964^{+0.0115}_{-0.0094}$
by Cudell et al. \cite{cudell}.

The final step in the analysis of the BPC data is to combine the 
$Q^2$ dependence from the GVMD fit with the energy dependence from
the Regge model
\begin{equation}
\sigma_{\rm tot}^{\gamma^* p}(W^2,Q^2) = 
\left({M_0^2\over M_0^2+Q^2}\right)
(A_R (W^2)^{\alpha_R-1}+ A_P(W^2)^{\alpha_P-1}).
\label{eqn:dl}
\end{equation}
The parameter $M_0^2$ is fixed to its value of $0.53$ found above and
$\alpha_R$ is also kept fixed at 0.5 as before. The 3 remaining
parameters are determined by fitting to photoproduction data (with 
$W^2 > 3\unit{GeV}^2$, but without the two original HERA measurements
) and the measured BPC data. We find
$A_R=145.0\pm 2.0\,\mu$b, $A_P=63.5\pm 0.9\,\mu$b and 
$\alpha_P=1.097\pm 0.002$
with $\chi^2/ndf=1.12$ (statistical errors only). If the two HERA 
photoproduction measurements are included
the parameters do not change within their errors. The description
of the low $Q^2$ $F_2$ data given by this model (ZEUSREGGE)
 is shown in 
\Ab{fig:dlfit}. Data in the BPC region $Q^2\leq 0.65\unit{GeV}^2$ are well
described. At larger $Q^2$ values the curves fall below the data.
Including ZEUS SVX95 data at successively larger $Q^2$ values, we find
that by $Q^2=3.5\unit{GeV}^2$ the $\chi^2/ndf$ has increased to
1.7. Also shown in \Ab{fig:dlfit}, for $Q^2\geq 0.9\unit{GeV}^2$, 
are the results of a NLO QCD fit (ZEUSQCD) described in \Se{sec:qcd}. 

To summarise, we have shown that the $Q^2$ dependence of the
ZEUS BPC95 data at very low $Q^2$ can be described by a simple
GVMD form. The resulting values of $\sigma_{0}^{\gamma p}$,
the cross-sections extrapolated to $Q^2 = 0$, are somewhat larger than 
the direct measurements at HERA. A two component Regge model gives a good 
description of the $W^2$ dependence of the data, with a Pomeron intercept 
compatible with that determined from hadron-hadron data. 

\section{$F_2$ slopes}
\label{sec:results}

To quantify the low $x$ behaviour of $F_2$ as a function of $x$ and $Q^2$,
we calculate the slopes $d\ln F_2/d\ln(1/x)$ at fixed $Q^2$ and 
$dF_2/d\ln Q^2$ at fixed $x$ from the ZEUS SVX95, BPC95 and ZEUS94 data sets.
We use the ALLM97 parameterisation \cite{allm97} for  
bin-centering $F_2$ data when necessary. 

\subsection{The slope $d\ln F_2/d\ln(1/x)$}
\label{sec:lameff}

It is seen from \Ab{fig:dlfit} that the $x$ slope of $F_2$ is small for 
small $Q^2$ and then starts to increase as $Q^2$ increases.
At a fixed value of $Q^2$ and at small $x$ the behaviour of
$F_2$ can be characterised by $F_2 \propto x^{-\lambda_{eff}}$
(giving $\lambda_{eff}=d\ln F_2/d\ln(1/x)$), with $\lambda_{eff}$ taking 
rather different values in the Regge and LO BFKL approaches. The value
of $\lambda_{eff}$ as an observable at small $x$ has been
discussed by  Navelet et al. \cite{nav1,nav2} and data on $\lambda_{eff}$ 
with $x<0.1$ have been presented by H1 \cite{h194}.

Using statistical errors only, we fit $F_2$ data at fixed $Q^2$ and $x<0.01$ 
to the form $A x^{-\lambda_{eff}}$. Refering to \Ab{fig:kinreg}, we are 
measuring $\lambda_{eff}$ from horizontal slices of $F_2$ data between 
the $y=1$ HERA kinematic limit and the fixed cut of $x<0.01$. As the 
$x$ range of the ZEUS BPC95 data is
restricted we include data from E665 \cite{E665F2}. In each $Q^2$ bin 
the average value of $x$, $\langle x \rangle$, is calculated from the 
mean value of $\ln(1/x)$ weighted 
by the statistical errors of the corresponding $F_2$ values in that bin. 
For the estimation of the systematic errors, it is
assumed that the systematic error analyses for each of the four
data sets used are independent. For a particular data set and a given
systematic check, the $F_2$
points in each $Q^2$ bin are moved up and down by the respective
systematic error estimates and the fits repeated, keeping all other
data sets fixed at their nominal values. The positive and negative shifts 
with respect to the central values of $\lambda_{eff}$ are added separately 
in quadrature to give the positive and negative systematic errors.

\Ab{fig:lambda_eff} shows the measured values of $\lambda_{eff}$ 
as a function of $Q^2$, and the data are given in \Ta{tab:lam} of the Appendix.
At very low $Q^2$ the errors on  $\lambda_{eff}$ are large because
this region is below the lower limit of E665 data (see \Ab{fig:kinreg}). 
At $Q^2>100\unit{GeV}^2$ the statistical error dominates. From the 
Regge approach of the previous
section one would expect $\lambda_{eff}\approx 0.1$, independently of
$Q^2$. Data for $Q^2<1 \unit{GeV}^2$ are consistent with this expectation.
The points labelled DL, linked by a dashed line, are calculated from the 
Donnachie-Landshoff prediction
\cite{dltwo} and as expected from the discussion of the previous
section are somewhat below the data. The variation of the DL points with
$Q^2$ is a consequence of averaging the model in a $Q^2$ bin over
a variable range of $x$ and hence $W^2$. 
For $Q^2>1\unit{GeV}^2$, $\lambda_{eff}$ increases to 
around $0.3$ at $Q^2$ values of $40\unit{GeV}^2$. Qualitatively the 
tendency of $\lambda_{eff}$ to increase with $Q^2$ is described by a 
number of pQCD approaches~\cite{nav2}. The points labelled GRV94, linked by 
a dashed line, are calculated from the NLO QCD GRV94 fit. Although the 
GRV94 prediction 
follows the trend of the data it tends to lie above the data, particularly 
in the $Q^2$ range $3-20\unit{GeV}^2$. We shall return to this point later.
For the predictions shown in \Ab{fig:lambda_eff} the same $F_2$ error 
weighted average in $x$ at a given $Q^2$ is used as for the data.

\subsection{The slope $dF_2/d\ln Q^2$}
\label{sec:df2dln2}

In QCD the scaling violations of $F_2$ are caused by 
gluon bremsstrahlung from quarks and quark pair creation from gluons. In 
the low $x$ domain accessible at HERA the latter process dominates the 
scaling violations. $F_2$ is then largely determined by the sea quarks 
$F_2\sim xS$, whereas the $dF_2/d\ln Q^2$ is dominated 
by the convolution of the splitting function $P_{qg}$ and the gluon density: 
$dF_2/d\ln Q^2\sim \alpha_S P_{qg}\otimes xg$. This has been used by 
Prytz~\cite{prytz} to relate $xg$ directly to the measured values of 
$dF_2/d\ln Q^2$~\cite{zg93}. The importance of $d F_2/d\ln Q^2$ as a tool 
for studying the low $x$ region was pointed out by Bartels et al.~\cite{bcf}.
 
In order to study the scaling violations of $F_2$ in more
detail the logarithmic slope $d F_2/d\ln Q^2$ is derived from the
data by fitting $F_2 = a+b\ln{Q^2}$ in bins of fixed $x$,
using only statistical errors. The ZEUS data sets used are the BPC95, 
SVX95 and ZEUS94. For compatibility with our NLO QCD fit a cut of 
$W^2>10\unit{GeV}^2$ is applied to the data. The $F_2$ data are shown
in bins of $x$ as functions of $Q^2$ in \Ab{fig:f2vq_binx}. The
fits $F_2=a+b\ln Q^2$ are also shown and the change of slope
as $x$ changes is visible from the plots.  In each $x$ bin the
average value of $Q^2$, $\langle Q^2 \rangle$,  is derived from the 
$F_2$ statistical error weighted mean value of $\ln Q^2$ in that bin. 
Systematic errors
on $d F_2/d\ln Q^2$ are estimated following the procedure outlined
in the previous section. The results for $d F_2/d\ln Q^2$ as a function of 
$x$ are shown in \Ab{fig:df2dlnq2} and are given in \Ta{tab:df} of the 
Appendix. The differences in the sizes of the errors on $d F_2/d\ln Q^2$ 
partially reflect the variations in $Q^2$ range as $x$ varies
(see \Ab{fig:kinreg} - we are taking vertical slices of the data to 
determine $d F_2/d\ln Q^2$). 
For values of $x$ down to $3\times 10^{-4}$, the slopes are increasing 
as $x$ decreases. At lower values of $x$ and $Q^2$, the slopes decreases.
If $dF_2/d\ln Q^2$ values are plotted for fixed target data at similar values
of $Q^2$, the `turn over' is not seen, but the data are at larger values 
of $x$~\cite{caldwell,mrst}. The points linked by the dashed lines are again 
from the DL Regge model and the GRV94 QCD fit, in both cases 
calculated using the same 
$F_2$ error weighted $Q^2$ averaging as for the data. The failure of 
DL is in line with our earlier discussion but the
fact that GRV94 does not follow the trend of the data when it turns over 
is perhaps more surprising. Naively it appears that the GRV94 gluon
density is too large even at $Q^2$ values around $5\unit{GeV}^2$.
We shall return to this discussion after we 
have presented the ZEUS NLO QCD fit to which we now turn.

\section{NLO QCD fit to $F_2$ data and extraction of
the gluon momentum density}
\label{sec:qcd}

In this section we present a NLO QCD fit to the ZEUS94
data~\cite{ZF2} and the SVX95 data of this paper.
We do not attempt a global fit to parton densities, but 
concentrate on what ZEUS data allow us to conclude
about the behaviour of the gluon and quark densities at low $x$.

To constrain the fits at high $x$, proton and deuteron
$F_2$ structure function data from NMC~\cite{mbb:newnmc} and
BCDMS~\cite{mbb:bcdms} are included.\footnote{Data from E665 are not
included in the fit. They are important at low $x$ and 
$Q^2<1\unit{GeV}^2$ but of much lower statistical weight at larger
$x$ compared to BCDMS and NMC. We have checked that including E665
data within the cuts described does not change the nominal fit result.}
 The following cuts are made on
the ZEUS and the fixed target data: (i)~$W^2 > 10$~\gev\ to reduce the
sensitivity to target mass~\cite{mbb:georgi} and higher 
twist~\cite{mbb:marcv} contributions which become important at high $x$
and low $Q^2$; (ii)~discard deuteron data with $x > 0.7$ to eliminate
possible contributions from Fermi motion in deuterium~\cite{mbb:bodek}. 
The kinematic range covered by the data input to the QCD fit 
is $3 \times 10^{-5} < x < 0.7$ and $ 1 < Q^2 < 5000$~\gev.

The QCD predictions for the $F_2$ structure functions
are obtained by solving the DGLAP evolution 
equations~\cite{dglap} at NLO in the \msbar\ scheme~\cite{mbb:furm}.
These equations yield the quark and gluon momentum distributions
(and thus the structure functions) at all values of $Q^2$ provided they
are given as functions of $x$ at some input scale $Q^2_0$.
In this analysis we adopt the so-called fixed flavour number scheme
where only three light flavours ($u,d,s$) contribute to the quark
density in the proton. In this scheme the assumption is made that
the charm and bottom quarks are produced in the hard scattering
process and the corresponding structure functions $F_2^c$ and
$F_2^b$ are calculated from the photon-gluon
fusion process including NLO corrections~\cite{mbb:f2c}.

As will be explained later, the input scale is chosen to be 
$Q^2_0 = 7$~GeV$^2$. The gluon distribution ($xg$),
the sea quark distribution ($xS$) and the difference of
up and down quarks in the proton ($x\Delta_{ud}$) are
parameterised as
\begin{eqnarray} \label{mbb:param}
xg(x,Q_0^2) & = & A_g x^{\de_g}(1-x)^{\eta_g} 
(1 + \gamma_g x) \nonumber \\
xS(x,Q_0^2) & \equiv &  2x ( \bar{u} + \bar{d} + \bar{s} ) =
A_s x^{\de_s}(1-x)^{\eta_s}(1+\varepsilon_s \sqrt{x}
 + \gamma_s x) \\
x\Delta_{ud}(x,Q_0^2) & \equiv & x(u+\bar{u}) - x(d+\bar{d}) =
A_{\Delta} x^{\delta_{\Delta}}(1-x)^{\eta_{\Delta}}.
  \nonumber 
\end{eqnarray}
The input valence distributions $xu_v = x(u-\bar{u})$ and 
$xd_v = x(d-\bar{d})$
at $Q^2_0$ are taken from the parton distribution set
MRS(R2)~\cite{mrsr}. As for MRS(R2) we assume that the 
strange quark distribution is a given fraction $K_s = 0.2$ of the sea
at the scale $Q^2 = 1$~\gev. The gluon normalisation, $A_g$, is fixed by the
momentum sum rule, \break
$\displaystyle \int^1_0(xg+xS+xu_v+xd_v)dx=1$.
There are thus 11 free parameters in the fit. 

The input value for the strong coupling constant is set to
$\asmz =  0.118$~\cite{mbb:pdg}. With a charm (bottom) threshold
of $Q_{c(b)} = 1.5$ (5)~GeV this corresponds to values of the QCD
scale parameter $\La_{\msbar} = (404,343,243)$~MeV for
$f = (3,4,5)$ flavours.
In the calculation of the charm
structure function $F_2^c$ the charm mass is taken to be
$m_c = 1.5$~GeV; the contribution from bottom is found to be negligible
in the kinematic range covered by the data. In the QCD evolutions and
the evaluation of the structure functions the renormalisation scale and
the mass factorisation scale are both set equal to $Q^2$.

The QCD evolutions and the structure function calculations are done
with the program QCDNUM~\cite{mbb:qcdnum}. The QCD evolution
equations are written in terms of quark flavour singlet and non-singlet
and gluon momentum distributions. The quark non-singlet evolution is
independent of the gluon. The quark singlet distribution
is defined as the sum over all quark and anti-quark distributions
\begin{equation}
x\Sigma=\sum_{i=u,d,s}[xq_i(x)+x\bar{q}_i(x)]
\end{equation}
and its evolution in $Q^2$ is coupled to that of the gluon distribution.
At small values of $x$, $x\Sigma$ is dominated by the contribution from
the $q\bar{q}$ sea $xS$. Note that for data
with $Q^2<Q^2_0=7\unit{GeV}^2$, backwards evolutions in $Q^2$ are
performed. The $\chi^2$ minimisation and
the calculation of the covariance matrices are based on
MINUIT~\cite{mbb:minuit}. In the definition of the $\chi^2$ only
statistical errors are included and the relative normalisations of the
data sets are fixed at unity.

The fit yields a good description of the data as shown in
\Ab{fig:f2q2fit} where we plot the $Q^2$ dependence of the proton 
structure function $F_2$ in the $x$ range covered by the ZEUS94 
data. The characteristic pattern of scaling violations can be seen
clearly from this plot, with $F_2$ at low values of $x$ rising as $Q^2$ 
increases. The quality of the fit to ZEUS data at
low $Q^2$ is also shown by the full line in \Ab{fig:svtx1}.
Adding the statistical and systematic errors in quadrature
gives a $\chi^2$ of 1474 for 1120 data points and 11 free parameters.
We have also checked that the gluon obtained from this fit to scaling
violations gives values of $F_2^c$ in agreement with the ZEUS 
measurements~\cite{zeusd*}. The values of the fitted parameters are given 
in \Ta{tab:qcd} of the Appendix.

\Ab{fig:xg1720} shows the gluon momentum distribution
as a function of $x$ for $Q^2$ at 1, 7 and $20\unit{GeV}^2$. 
The following sources contribute to the inner shaded error
bands displayed in the figure (for each source we give in brackets
the relative error $\Delta g / g$ at $x = 5 \times 10^{-5}$,
$Q^2 = 7$~\gev):
\begin{enumerate}
\item 
The statistical error on the data (4\%).
\item 
The experimental systematic errors, (13\%), which are propagated to the
covariance matrix of the fitted parameters using the technique described
in~\cite{mbb:syserr}. In total 26 independent sources of systematic error
are included. Proper account is taken of the correlations between the
systematic errors of the NMC datasets and for BCDMS the procedure of
\cite{mbb:marcv} is followed. Normalisation errors of all data sets are also 
included.
\item 
The uncertainties on the strong coupling constant 
$\Delta\alpha_s(M_Z^2)=0.005$ (8\%),
on the strange quark content of the proton $\Delta K_s = 0.1$ (1\%) and on
the charm mass $\Delta m_c = 0.2$ GeV (1\%).
\end{enumerate}
Adding errors from (1), (2) and (3) together in quadrature gives a total 
contribution of 16\%.\footnote{This combination of errors 
(`HERA standard errors') is often used by the H1 and ZEUS experiments when 
discussing $xg$.} 

In addition to the above sources of error a `parameterisation error' (10\%) 
is obtained by repeating the fit with:
\begin{enumerate}
  \setcounter{enumi}{3}
  \item 
  The addition of statistical and systematic errors in quadrature in the
  definition of the $\chi^2$ instead of taking statistical
  errors only.
  \item 
   The input scale set to $Q^2_0 = 1$ and 4~\gev\ instead of 7~\gev.
  \item 
   An alternative parameterisation of the gluon density:
  \begin{equation} \label{mbb:cheb}
   xg(x,Q^2_0) = A_g (1-x)^{\eta_g} \left[ 1 + \sum_{n=1}^3 C_n T_n(y) \right]
  \end{equation}
  where $T_n(y)$ is a Chebycheff polynomial of the first kind~\cite{mbb:stegun}
  and $y = a \ln x + b$ with the coefficients $a,b$ adjusted such that
  $x \in [10^{-6},1]$ maps onto $y \in [-1,1]$. This parameterisation is 
  flexible enough to describe the rapid change with $Q^2$ (see \Ab{fig:xg1720})
  of the shape of the gluon density. Furthermore, \Gl{mbb:cheb} allows the 
  gluon density to become negative at low $x$ whereas \Gl{mbb:param} imposes 
  the constraint $xg(x)\geq 0$ as $x \to 0$.
\end{enumerate}
Taking all combinations of the alternatives described in 4., 5. and
6. above in addition to the nominal settings,
twelve fits are performed and the parameterisation error
is defined as the envelope of the resulting set of quark and gluon
distributions. All these fits yield similar values of $\chi^2$. 
The nominal fit (ZEUSQCD) is taken to be that which gives a curve
that is roughly at the centre of the error bands for all $x$ and $Q^2$.
This also defines the choice of $Q^2_0=7\unit{GeV}^2$.
The outer hatched error bands in \Ab{fig:xg1720} correspond 
to the total error now including the parameterisation error added in
quadrature with the other errors. At 
$x = 5 \times 10^{-5}$, $Q^2 = 7$~\gev ~the total $\Delta g / g = 19$\%.

The three left-hand plots of \Ab{fig:xSxg} show the distributions for 
$x\Sigma$ and $xg$ as functions of $x$ for $Q^2$ at 1, 7 and $20\unit{GeV}^2$.
The error bands shown correspond to the quadratic sum of all error
sources. It can be seen that even at the smallest $Q^2$ the quark singlet
distribution, $x\Sigma$, is rising at small $x$ whereas the gluon 
distribution, xg, has become almost flat, indeed compatible with zero. 
This behaviour has also been found by others, for instance Martin et al. 
(MRST) \cite{mrst} in their recent global determination of parton densities.
At $Q^2 = 1$~\gev\ the gluon distribution is poorly
determined and can, within errors, be negative at low $x$. In the 
simplest form of the parton model (and leading order QCD)
this would clearly be unphysical
and while it is known that at NLO in the \msbar\ scheme a positive 
parton density will remain positive for forward evolution in $Q^2$
there is no such constraint for backwards evolution \cite{stirling}.
A negative gluon distribution is therefore not necessarily in contradiction 
with perturbative NLO QCD provided cross-sections or structure functions 
calculated from the parton distributions are positive for all $x$ and $Q^2$ 
in the fitted kinematic domain.  We have verified that
this is the case for $F_2^c$ and the longitudinal structure function
$F_L$.\footnote{Here and in the following $F_L$ is calculated using the
QCD calculation of Altarelli and Martinelli \cite{amfl}.}

\section{The transition region and NLO QCD}
\label{sec:loq2qcd}
\subsection{The NLO QCD fit at low $Q^2$}

It is now widely observed that NLO DGLAP QCD fits give good descriptions
of $F_2$ data down to $Q^2$ values in the range $1-2\unit{GeV}^2$.
For such fits to be valid one assumes:
\begin{enumerate}
\item
the validity of the DGLAP QCD formalism;
\item
NLO is sufficient even though $\alpha_S$ is becoming large (e.g. for this
analysis $\alpha_S=0.46$ at $1\unit{GeV}^2$ );
\item
that no higher twist terms, shadowing or other 
non-perturbative effects contribute to $F_2$.
\end{enumerate}
In this paper we have also deliberately made the minimum of assumptions
about the low $x$ functional form of the parton distributions at $Q^2_0$.
We require only that they must tend to zero as $x\to 1$ and that the 
flavour and momentum sum-rules are respected.

To investigate the stability of our results at low $Q^2$ we have repeated 
the full QCD fit and error evaluation
procedure on the same data as in \Se{sec:qcd} but with the minimum 
$Q^2$ cut ($Q^2_{min}$) raised to $4\unit{GeV}^2$. The quality of the fit 
is much as before,
$\chi^2$ of 1242 for 943 data points and 11 parameters (statistical
and systematic errors added in quadrature). The resulting $x\Sigma$ and
$xg$ parton distributions are shown in the three right-hand plots of 
\Ab{fig:xSxg}. Qualitatively the features shown by the standard fit (left-hand
plots) are unchanged. The rising $x\Sigma$ distribution at low $x$ remains 
and the sea dominates the gluon at small $x$ and the lowest $Q^2$ value. 
In more detail: 
\begin{itemize}
\item
The parton densities from the central fits with $Q^2_{min}=1$ and 
$4\unit{GeV}^2$ are very similar.
\item
Except at the lowest $Q^2$, the precision of the determination of $x\Sigma$ 
is not much reduced. Even at $Q^2=1\unit{GeV}^2$, $x\Sigma$ is reasonably 
well determined for $x>3\times 10^{-4}$, at smaller $x$ values there 
are insufficient data to constrain the fit.
\item
At all $Q^2$ values shown in the right-hand plots the precision of 
the determination of $xg$ for $x<10^{-3}$ is worse.
\end{itemize}
The increase in the
error bands for the gluon density when the $Q^2_{min}$ cut is increased shows
the importance of the SVX95 data at low $x$ and low $Q^2$ in
determining $xg$ in this region.

To investigate if there is a technical lower limit to the NLO QCD fit
(in the sense that the fit fails to converge or gives a very bad $\chi^2$),
we extend the QCD fit into the region covered by the ZEUS BPC95 
data by lowering the $Q^2_{min}$ cut to 0.4~GeV$^2$. 
The fit gives an acceptable description of the data with $F_2^c$
positive and $F_L$ (calculated as in \cite{amfl}) only slightly
negative at $x \approx 5 \times 10^{-3}$ and $Q^2 = 0.5$~GeV$^2$.
We therefore conclude that we do not observe a significant breakdown
of the NLO DGLAP description in the kinematic range explored.
However, we stress that, within the present experimental accuracy,
$F_2$ data by itself cannot validate the assumptions noted at the beginning
of this section and that other information such as precise measurements
of $F_2^c$ in DIS, measurements of other hard processes or more 
theoretical input is required.~\footnote{Some of the features that 
we find, such as the suppression of the gluon density at small $Q^2$, and 
the ability of the NLO DGLAP formalism to fit very low $Q^2$ data have 
been noted previously by Lopez et al.~\cite{lopez}.}

\subsection{$F_2$ slopes and models}

In order to clarify what we have learned about the transition region
from the ZEUS Regge fit and the NLO DGLAP fits we return to the $F_2$ slopes.

\Ab{fig:lambda_eff}, showing the ZEUS+E665 $\lambda_{eff}$ data of
\Se{sec:lameff}, also shows the calculation
from the ZEUSREGGE fit of \Se{sec:lowq2reg} and for $Q^2>1\unit{GeV}^2$ 
the result from the  ZEUSQCD fit of \Se{sec:qcd}. 
In the BPC region, the ZEUSREGGE (full line)
calculation gives $\lambda_{eff}$ somewhat higher than that given
by the original DL Regge fit (dashed line), but it is 
still below the data. This is largely because the ZEUSREGGE fit 
includes low $W$ photoproduction data and thus gives a lower
$\alpha_P$ than the value of $1.141$ from the Regge fit to the extrapolated
ZEUS BPC $\sigma_0^{\gamma p}$ data alone. At larger $Q^2$ values, 
ZEUSQCD (full line) gives a good account of the trend and normalisation of  
$\lambda_{eff}$ while GRV94 (dashed line) tends to predict a larger value of
$\lambda_{eff}$, which means a steeper rise of $F_2$ as $x$ decreases,
than that determined from the data. 

\Ab{fig:df2dlnq2} shows the ZEUS $dF_2/d\ln Q^2$ data
together with the same two calculations, ZEUSREGGE at very low $x$ and 
$Q^2$ values
and ZEUSQCD for $Q^2>1\unit{GeV}^2$. The ZEUSREGGE points show much the 
same trend as that of the original DL model.
At $Q^2$ values between 1 and $5\unit{GeV}^2$, the ZEUSQCD points are 
qualitatively different from those of GRV94. The ZEUSQCD values
now follow the `turn over' of the slope data around $x\sim3\times 10^{-4}$.
This has also been found by MRST in \cite{mrst} where they compare their 
global fit to $dF_2/d\ln Q^2$ data from H1, ZEUS and the NMC 
experiments. The rise in $F_2$ at low $x$ and the `turn over' in 
$dF_2/d\ln Q^2$ reflect the different behaviours of the gluon and 
$q\bar{q}$ sea distributions at $Q^2\sim 1\unit{GeV}^2$. For very small 
$x$ the sea continues to rise ($\delta_s<0$ in the notation of \Gl{mbb:param}) 
whereas the gluon rises significantly less steeply or even tends 
to zero ($\delta_g\geq 0$).

In contrast, for the GRV94 parton densities GRV assume that at their 
low starting scale of 
$0.34\unit{GeV}^2$ {\it both} the gluon and the sea distributions are 
non-singular. All the rise of $F_2$ at low $x$ for $Q^2>1\unit{GeV}^2$ is 
then generated through the DGLAP evolution equations.
In a recent paper GRV~\cite{grv98} have revisited their `dynamical 
parton model' in the light of the HERA 1994 data and have produced 
a new set of parton distributions -- GRV98. They find that they can
correct most of the discrepancies between GRV94 and the HERA data
by a slight increase in the starting scale from $0.34$ to 
$0.4\unit{GeV}^2$ and by using a lower value of $\alpha_S(M^2_Z)=0.114$ 
rather than the value of $0.118$ used in the ZEUS NLO QCD fit. They 
acknowledge that if they use a larger value of $\alpha_S$ then their starting
scale has to be increased to around $1\unit{GeV}^2$ and they
have to accept a rising $q\bar{q}$ sea distribution at the starting scale. 

\section{Summary and conclusions}
\label{sec:concl}

In this paper we have presented the measurement by ZEUS of $F_2$ in the
$Q^2$ region $0.6 - 17\unit{GeV}^2$ (SVX95), which fills the gap 
between the very low $Q^2$ BPC95 data ($0.11 - 0.65\unit{GeV}^2$) 
and the large 1994 data sample ($3.5 - 5000\unit{GeV}^2$). 
We have shown that the BPC data may be described 
successfully by non-perturbative approaches: a simple generalised vector 
dominance model for the $Q^2$ dependence and a two component Regge 
model for the $W^2$ dependence. For $Q^2 \geq 0.9\unit{GeV}^2$ these
approaches fail to describe the dominant feature of the data, which
is the rapid rise of $F_2$ at small $x$. 

We have studied the transition region by fitting 
$F_2=Ax^{-\lambda_{eff}}$ using ZEUS and E665 data with $x<0.01$ in
the $Q^2$ range $0.15 - 250\unit{GeV}^2$. For $Q^2>0.9\unit{GeV}^2$
the data are not compatible with the $Q^2$ independence of 
$\lambda_{eff}$, as expected for a dominant Pomeron term
in conventional Regge theory, but are well described by the ZEUS NLO QCD fit. 

The slope $dF_2/d\ln Q^2$ has been calculated from ZEUS $F_2$ data in the
range $2\times 10^{-6}<x<0.2$. Assuming pQCD, $F_2$ at low $x$ is 
largely determined by the $q\bar{q}$ sea density whereas $dF_2/d\ln Q^2$ is 
given by the gluon density. As $x$ decreases the slope values increase 
until at $x\approx 3\times 10^{-4}$ there is a turn over and
for smaller $x$ values the slope values decrease.

To study the behaviour of the parton densities in more detail we have
performed a NLO DGLAP QCD fit to ZEUS and fixed target data with
$Q^2>1\unit{GeV}^2$ and $3\times10^{-5}<x<0.7$. A good description of
the $F_2$ data over the whole range of $Q^2$ from 1 to $5000\unit{GeV}^2$
is obtained. Around the lower $Q^2$ limit of the fit we find that the 
$q\bar{q}$ sea distribution is still
rising at small $x$, whereas the gluon distribution is strongly
suppressed. These findings are incompatible with the 
hypothesis that the rapid rise in $F_2$ is driven by the rapid increase in 
gluon density at small $x$ from parton splitting alone. 
These features remain true if the lowest $Q^2$ for 
which data is included in the fit is raised from 1 to $4\unit{GeV}^2$. 

From the ZEUS QCD fit we also obtain a much improved 
determination of the gluon 
momentum density compared to the previous determination 
by ZEUS \cite{zg93}. Full account has been taken of correlated experimental 
systematic errors and the uncertainty in the form of the input gluon 
distribution function has also been estimated. At $Q^2=20\unit{GeV}^2$ and 
$x=5\times 10^{-5}$ the total fractional error on the gluon density has been 
reduced from 40\% to 10\%. 

We have used NLO pQCD and the simplest non-perturbative
models to study the transition region in $Q^2$. We find,
for $Q^2\geq 0.9\unit{GeV}^2$, that the wholly non-perturbative 
description fails. Although pQCD may not be valid at such low scales, we 
have not been able to find a lower limit at which the NLO DGLAP
fit breaks down conclusively.

\section*{Acknowledgements}

The experiment was made possible by the skill and dedication of the
HERA machine group who ran HERA most efficiently during 1994 and 1995
when the data for this paper were collected. The realisation and
continuing operation of the ZEUS detector has been and is made possible by
the inventiveness and continuing hard work of many people not listed as
authors. Their contributions are acknowledged with great appreciation.
The support and encouragement of the DESY directorate continues to be
invaluable for the successful operation of the ZEUS collaboration. 
We thank J. Bl\"umlein, R. Thorne and W. J. Stirling for discussions
on QCD.


\section*{Appendix}
\bigskip
\par\noindent
The following tables are also available via the ZEUS collaboration
home page, \break http://www-zeus.desy.de/.
Fortran routines and data files to calculate the parton distributions
from the ZEUS NLO QCD fits are also available from this site.

\begin{table}
\begin{center}
\begin{tabular}{|c|c|c|c|c|c|} %
\hline
 bin & $x$ &  $Q^2$ ($\unit{GeV}^2$) & $F_2$ & error $ stat._{-sys}^{+sys}$ & 
 $F_2~(F_L = 0)$\\
\hline
\hline
1 &$  1.2\cdot 10^{-5}$&$  0.6$&$  0.531$&$\pm  0.030_{-0.060}^{+0.037}$ &$0.517  $\\
2 &$  1.9\cdot 10^{-5}$&$  0.9$&$  0.653$&$\pm  0.030_{-0.053}^{+0.067}$ &$0.632  $\\
3 &$  2.6\cdot 10^{-5}$&$  0.9$&$  0.655$&$\pm  0.023_{-0.037}^{+0.039}$ &$0.644  $\\
4 &$  3.6\cdot 10^{-5}$&$  0.9$&$  0.574$&$\pm  0.022_{-0.050}^{+0.034}$ &$0.570  $\\
5 &$  2.8\cdot 10^{-5}$&$  1.3$&$  0.716$&$\pm  0.029_{-0.065}^{+0.069}$ &$0.687  $\\
6 &$  3.8\cdot 10^{-5}$&$  1.3$&$  0.717$&$\pm  0.023_{-0.037}^{+0.043}$ &$0.703  $\\
7 &$  5.3\cdot 10^{-5}$&$  1.3$&$  0.715$&$\pm  0.022_{-0.041}^{+0.036}$ &$0.708  $\\
8 &$  8.0\cdot 10^{-5}$&$  1.3$&$  0.626$&$\pm  0.017_{-0.031}^{+0.033}$ &$0.624  $\\
9 &$  1.6\cdot 10^{-4}$&$  1.3$&$  0.559$&$\pm  0.016_{-0.047}^{+0.046}$ &$0.558  $\\
10&$  3.9\cdot 10^{-5}$&$  1.9$&$  0.915$&$\pm  0.036_{-0.070}^{+0.071}$ &$0.873  $\\
11&$  5.3\cdot 10^{-5}$&$  1.9$&$  0.852$&$\pm  0.028_{-0.051}^{+0.033}$ &$0.833  $\\
12&$  7.5\cdot 10^{-5}$&$  1.9$&$  0.742$&$\pm  0.025_{-0.026}^{+0.038}$ &$0.734  $\\
13&$  1.1\cdot 10^{-4}$&$  1.9$&$  0.740$&$\pm  0.023_{-0.038}^{+0.035}$ &$0.736  $\\
14&$  2.2\cdot 10^{-4}$&$  1.9$&$  0.636$&$\pm  0.019_{-0.057}^{+0.026}$ &$0.635  $\\
15&$  5.4\cdot 10^{-5}$ &$  2.5$&$  0.964$&$\pm  0.039_{-0.060}^{+0.060}$ &$0.916  $\\
16&$  7.3\cdot 10^{-5} $&$  2.5$&$  0.909$&$\pm  0.029_{-0.040}^{+0.044}$ &$0.887  $\\
17&$  1.0\cdot 10^{-4} $&$  2.5$&$  0.870$&$\pm  0.027_{-0.044}^{+0.022}$ &$0.860  $\\
18&$  1.6\cdot 10^{-4}$&$  2.5$&$  0.771$&$\pm  0.023_{-0.016}^{+0.050}$ &$0.767  $\\
19&$  3.0\cdot 10^{-4}$&$  2.5$&$  0.725$&$\pm  0.023_{-0.060}^{+0.029}$ &$0.724  $\\
20&$  7.3\cdot 10^{-5}$&$  3.5$&$  1.140$&$\pm  0.050_{-0.057}^{+0.058}$ &$1.081  $\\
21&$  9.8\cdot 10^{-5}$&$  3.5$&$  0.995$&$\pm  0.035_{-0.023}^{+0.040}$ &$0.970  $\\
22&$  1.4\cdot 10^{-4}$&$  3.5$&$  0.945$&$\pm  0.031_{-0.022}^{+0.034}$ &$0.933  $\\
23&$  2.1\cdot 10^{-4}$&$  3.5$&$  0.839$&$\pm  0.026_{-0.028}^{+0.019}$ &$0.834  $\\
24&$  4.1\cdot 10^{-4}$&$  3.5$&$  0.663$&$\pm  0.021_{-0.048}^{+0.044}$ &$0.661  $\\
25&$  1.0\cdot 10^{-4}$&$  4.5$&$  1.160$&$\pm  0.043_{-0.044}^{+0.040}$ &$1.108  $\\
26&$  1.7\cdot 10^{-4}$&$  4.5$&$  1.005$&$\pm  0.031_{-0.037}^{+0.050}$ &$0.990  $\\
27&$  2.8\cdot 10^{-4}$&$  4.5$&$  0.912$&$\pm  0.030_{-0.032}^{+0.029}$ &$0.907  $\\
28&$  5.4\cdot 10^{-4}$&$  4.5$&$  0.664$&$\pm  0.022_{-0.057}^{+0.055}$ &$0.663  $\\
29&$  1.4\cdot 10^{-4}$&$  6$&$  1.305$&$\pm  0.055_{-0.037}^{+0.041}$ &$1.247  $\\
30&$  2.2\cdot 10^{-4}$&$  6$&$  1.086$&$\pm  0.039_{-0.046}^{+0.025}$ &$1.070  $\\
31&$  3.7\cdot 10^{-4}$&$  6$&$  1.002$&$\pm  0.038_{-0.033}^{+0.027}$ &$0.996  $\\
32&$  7.1\cdot 10^{-4}$&$  6$&$  0.753$&$\pm  0.028_{-0.079}^{+0.050}$ &$0.751  $\\
33&$  1.8\cdot 10^{-4}$&$  7.5$&$  1.210$&$\pm  0.058_{-0.051}^{+0.037}$ &$1.167  $\\
34&$  3.6\cdot 10^{-4}$&$  7.5$&$  0.918$&$\pm  0.036_{-0.023}^{+0.074}$ &$0.910  $\\
35&$  8.8\cdot 10^{-4}$&$  7.5$&$  0.810$&$\pm  0.040_{-0.038}^{+0.060}$ &$0.808  $\\
36&$  2.2\cdot 10^{-4}$&$  9$&$  1.296$&$\pm  0.061_{-0.074}^{+0.039}$ &$1.251  $\\
37&$  4.3\cdot 10^{-4}$&$  9$&$  0.957$&$\pm  0.038_{-0.036}^{+0.028}$ &$0.948  $\\
38&$  1.1\cdot 10^{-3}$&$  9$&$  0.818$&$\pm  0.039_{-0.065}^{+0.027}$ &$0.816  $\\
39&$  2.8\cdot 10^{-4}$&$ 12$&$  1.306$&$\pm  0.057_{-0.038}^{+0.066}$ &$1.265  $\\
40&$  5.6\cdot 10^{-4}$&$ 12$&$  1.084$&$\pm  0.040_{-0.061}^{+0.043}$ &$1.073  $\\
41&$  1.4\cdot 10^{-3}$&$ 12$&$  0.956$&$\pm  0.044_{-0.040}^{+0.057}$ &$0.954  $\\
42&$  3.9\cdot 10^{-4}$&$ 17$&$  1.311$&$\pm  0.057_{-0.055}^{+0.040}$ &$1.270  $\\
43&$  7.9\cdot 10^{-4}$&$ 17$&$  1.087$&$\pm  0.038_{-0.041}^{+0.051}$ &$1.077  $\\
44&$  1.9\cdot 10^{-3}$&$ 17$&$  0.931$&$\pm  0.045_{-0.077}^{+0.046}$ &$0.927  $\\
\hline
\end{tabular}
\end{center}
\caption{Values of $x$, $Q^2$, $F_2$, statistical and systematic errors
from the ZEUS 1995 shifted vertex analysis (SVX95).
\label{tab:svx95}}
\end{table}

\begin{table}
\begin{center}
\begin{tabular}[htp]{|c|c|c|c|c|c|} %
\hline
bin &  $W$  &  $\epsilon$  & $\sigma^{\gamma p}_{0}$  &  
  stat. err. & sys. err.\\
 & [GeV]& & [$\mu$b] &[$\mu$b] &[$\mu$b] \\
\hline
\hline
 1& 104 & 0.99 & 156.2 & $\pm 5.3$ & $\pm 16.1$ \\
 2& 134 & 0.98 & 166.1 & $\pm 5.2$ & $\pm 11.0$ \\
 3& 153 & 0.96 & 174.7 & $\pm 4.9$ & $\pm 12.9$ \\
 4& 173 & 0.92 & 175.5 & $\pm 5.0$ & $\pm 11.7$ \\
 5& 190 & 0.88 & 181.8 & $\pm 4.7$ & $\pm 12.8$ \\
 6& 212 & 0.80 & 186.8 & $\pm 4.8$ & $\pm 13.5$ \\
 7& 233 & 0.69 & 192.5 & $\pm 4.7$ & $\pm 13.3$ \\
 8& 251 & 0.55 & 204.8 & $\pm 5.6$ & $\pm 17.0$ \\
\hline
\end{tabular}
\end{center}
\caption{Values of $\sigma^{\gamma p}_{0}$ together with statistical
and systematic errors from the GVMD extrapolation, with $\sigma_L=0$, 
of ZEUS BPC95 data. $W$ is the $\gamma^*p$ centre-of-mass energy and
$\epsilon$ is the polarisation of the virtual photon.
\label{tab:sigt}
}
\end{table}

\begin{table}
\begin{center}
\begin{tabular}[htp]{|c|c|c|c|c|c|c|} %
\hline
bin &  $Q^2$ ($\unit{GeV}^2$) &  $x_{min}$  & $x_{max}$  &   
$\langle x\rangle$ & 
$\lambda_{eff}$ & error $ stat.^{+sys}_{-sys}$\\
\hline
\hline
 1 &   0.15 & 2.4$\cdot 10^{-6}$ & 4.2$\cdot 10^{-6}$ & 3.2$\cdot 10^{-6}$ & 0.134 & $\pm 0.077^{+0.189}_{-0.194}$\\
 2 &   0.2 & 3.2$\cdot 10^{-6}$ & 8.5$\cdot 10^{-6}$ & 5.5$\cdot 10^{-6}$ & 0.128 & $\pm 0.036^{+0.080}_{-0.083}$\\
 3 &   0.25 & 4.6$\cdot 10^{-6}$ & 1.8$\cdot 10^{-3}$ & 1.3$\cdot 10^{-5}$ & 0.136 & $\pm 0.016^{+0.040}_{-0.044}$\\
 4 &   0.3 & 6.7$\cdot 10^{-6}$ & 2.5$\cdot 10^{-3}$ & 1.1$\cdot 10^{-4}$ & 0.110 & $\pm 0.007^{+0.017}_{-0.022}$\\
 5 &   0.4 & 1.1$\cdot 10^{-5}$ & 5.2$\cdot 10^{-3}$ & 6.8$\cdot 10^{-4}$ & 0.115 & $\pm 0.005^{+0.015}_{-0.019}$\\
 6 &   0.5 & 2.1$\cdot 10^{-5}$ & 4.6$\cdot 10^{-5}$ & 2.9$\cdot 10^{-5}$ & 0.193 & $\pm 0.050^{+0.100}_{-0.097}$\\
 7 &   0.65 & 1.2$\cdot 10^{-5}$ & 5.2$\cdot 10^{-3}$ & 1.8$\cdot 10^{-3}$ & 0.114 & $\pm 0.007^{+0.016}_{-0.019}$\\
 8 &   0.9 & 1.9$\cdot 10^{-5}$ & 8.9$\cdot 10^{-3}$ & 3.8$\cdot 10^{-3}$ & 0.147 & $\pm 0.005^{+0.013}_{-0.014}$\\
 9 &   1.3 & 2.8$\cdot 10^{-5}$ & 8.9$\cdot 10^{-3}$ & 2.4$\cdot 10^{-3}$ & 0.152 & $\pm 0.006^{+0.014}_{-0.014}$\\
10 &   1.9 & 3.9$\cdot 10^{-5}$ & 8.9$\cdot 10^{-3}$ & 1.8$\cdot 10^{-3}$ & 0.160 & $\pm 0.008^{+0.014}_{-0.013}$\\
11 &   2.5 & 5.4$\cdot 10^{-5}$ & 8.9$\cdot 10^{-3}$ & 1.3$\cdot 10^{-3}$ & 0.179 & $\pm 0.010^{+0.016}_{-0.014}$\\
12 &   3.5 & 6.3$\cdot 10^{-5} $ & 8.9$\cdot 10^{-3}$ & 6.1$\cdot 10^{-4}$ & 0.178 & $\pm 0.012^{+0.022}_{-0.018}$\\
13 &   4.5 & 1.0$\cdot 10^{-4}$ & 5.4$\cdot 10^{-4}$ & 2.0$\cdot 10^{-4}$ & 0.261 & $\pm 0.020^{+0.042}_{-0.041}$\\
14 &   6.5 & 1.0$\cdot 10^{-4}$ & 4.0$\cdot 10^{-3}$ & 7.9$\cdot 10^{-4}$ & 0.192 & $\pm 0.005^{+0.016}_{-0.012}$\\
15 &   8.5 & 1.6$\cdot 10^{-4}$ & 1.6$\cdot 10^{-3}$ & 6.7$\cdot 10^{-4}$ & 0.261 & $\pm 0.012^{+0.016}_{-0.020}$\\
16 &  10 & 1.6$\cdot 10^{-4}$ & 6.3$\cdot 10^{-3}$ & 1.6$\cdot 10^{-3}$ & 0.226 & $\pm 0.007^{+0.015}_{-0.014}$\\
17 &  12 & 2.5$\cdot 10^{-4}$ & 2.5$\cdot 10^{-3}$ & 9.7$\cdot 10^{-4}$ & 0.250 & $\pm 0.014^{+0.018}_{-0.020}$\\
18 &  15 & 2.5$\cdot 10^{-4}$ & 6.3$\cdot 10^{-3}$ & 1.9$\cdot 10^{-3}$ & 0.249 & $\pm 0.010^{+0.016}_{-0.014}$\\
19 &  18 & 3.9$\cdot 10^{-4}$ & 6.3$\cdot 10^{-3}$ & 2.1$\cdot 10^{-3}$ & 0.274 & $\pm 0.010^{+0.014}_{-0.013}$\\
20 &  22 & 4.0$\cdot 10^{-4}$ & 1.0$\cdot 10^{-2}$ & 3.4$\cdot 10^{-3}$ & 0.273 & $\pm 0.010^{+0.017}_{-0.014}$\\
21 &  27 & 6.3$\cdot 10^{-4}$ & 6.3$\cdot 10^{-3}$ & 2.7$\cdot 10^{-3}$ & 0.332 & $\pm 0.015^{+0.017}_{-0.014}$\\
22 &  35 & 6.3$\cdot 10^{-4}$ & 1.0$\cdot 10^{-2}$ & 3.7$\cdot 10^{-3}$ & 0.308 & $\pm 0.016^{+0.013}_{-0.019}$\\
23 &  45 & 1.0$\cdot 10^{-3}$ & 1.0$\cdot 10^{-2}$ & 3.7$\cdot 10^{-3}$ & 0.289 & $\pm 0.020^{+0.022}_{-0.017}$\\
24 &  60 & 1.0$\cdot 10^{-3}$ & 1.0$\cdot 10^{-2}$ & 4.7$\cdot 10^{-3}$ & 0.342 & $\pm 0.021^{+0.009}_{-0.017}$\\
25 &  70 & 1.6$\cdot 10^{-3}$ & 1.0$\cdot 10^{-2}$ & 4.8$\cdot 10^{-3}$ & 0.306 & $\pm 0.032^{+0.018}_{-0.023}$\\
26 &  90 & 1.6$\cdot 10^{-3}$ & 1.0$\cdot 10^{-2}$ & 5.3$\cdot 10^{-3}$ & 0.273 & $\pm 0.040^{+0.046}_{-0.027}$\\
27 & 120 & 2.5$\cdot 10^{-3}$ & 1.0$\cdot 10^{-2}$ & 6.2$\cdot 10^{-3}$ & 0.380 & $\pm 0.059^{+0.001}_{-0.002}$\\
28 & 150 & 2.5$\cdot 10^{-3}$ & 1.0$\cdot 10^{-2}$ & 6.3$\cdot 10^{-3}$ & 0.258 & $\pm 0.079^{+0.004}_{-0.004}$\\
29 & 200 & 4.0$\cdot 10^{-3}$ & 1.0$\cdot 10^{-2}$ & 7.0$\cdot 10^{-3}$ & 0.250 & $\pm 0.126^{+0.009}_{-0.003}$\\
30 & 250 & 4.0$\cdot 10^{-3}$ & 1.0$\cdot 10^{-2}$ & 7.9$\cdot 10^{-3}$ & 0.452 & $\pm 0.184^{+0.008}_{-0.000}$\\
\hline
\end{tabular}
\end{center}
\caption{Values of the slope $\lambda_{eff}=d\ln F_2/d\ln(1/x)$
and their errors, calculated 
from fitting $F_2=Ax^{-\lambda_{eff}}$ at fixed $Q^2$ to ZEUS and E665
data with $x<0.01$. The columns labelled $x_{min}, x_{max}$ and 
$\langle x\rangle$ give the minimum, maximum and
average values of $x$ in a bin. $\langle x\rangle$ is calculated as 
described in \Se{sec:lameff} of the text. 
\label{tab:lam}
}
\end{table}

\begin{table}
\begin{center}
\begin{tabular}[htp]{|c|c|c|c|c|c|c|} %
\hline
bin &  $x$ & $Q^2_{min}$ ($\unit{GeV}^2$)  & $Q^2_{max}$  &  
$\langle Q^2\rangle$ & 
$dF_2/d\ln Q^2$ & error $ stat.^{+sys}_{-sys}$\\
\hline
\hline
 1 & 2.1$\cdot 10^{-6}$ &  0.11 &     0.15 &  0.12 & 0.135 & $\pm 0.029^{+0.030}_{-0.029}$\\
 2 & 3.1$\cdot 10^{-6}$ &  0.15 &     0.2 &  0.16 & 0.198 & $\pm 0.028^{+0.033}_{-0.033}$\\
 3 & 4.6$\cdot 10^{-6}$ &  0.15 &     0.25 &  0.2 & 0.174 & $\pm 0.017^{+0.047}_{-0.047}$\\
 4 & 7.3$\cdot 10^{-6}$ &  0.2 &     0.3 &  0.23 & 0.191 & $\pm 0.019^{+0.040}_{-0.040}$\\
 5 & 1.2$\cdot 10^{-5}$ &  0.25 &     0.6 &  0.29 & 0.265 & $\pm 0.017^{+0.020}_{-0.023}$\\
 6 & 2.0$\cdot 10^{-5}$ &  0.3 &     0.9 &  0.37 & 0.297 & $\pm 0.018^{+0.026}_{-0.022}$\\
 7 & 3.3$\cdot 10^{-5}$ &  0.3 &     1.9 &  0.52 & 0.312 & $\pm 0.009^{+0.032}_{-0.031}$\\
 8 & 6.3$\cdot 10^{-5}$ &  0.5 &     3.5 &   1.1 & 0.365 & $\pm 0.010^{+0.032}_{-0.032}$\\
 9 & 1.0$\cdot 10^{-4}$ &  1.3 &     6.5 &   2.5 & 0.379 & $\pm 0.018^{+0.049}_{-0.043}$\\
10 & 1.6$\cdot 10^{-4}$ &  1.3 &    10 &   3.8 & 0.387 & $\pm 0.013^{+0.038}_{-0.038}$\\
11 & 2.5$\cdot 10^{-4}$ &  1.9 &    15 &   5.2 & 0.368 & $\pm 0.013^{+0.043}_{-0.029}$\\
12 & 4.0$\cdot 10^{-4}$ &  3.5 &    22 &   8.8 & 0.429 & $\pm 0.018^{+0.044}_{-0.035}$\\
13 & 6.3$\cdot 10^{-4}$ &  4.5 &    35 &    10 & 0.404 & $\pm 0.015^{+0.040}_{-0.035}$\\
14 & 1.0$\cdot 10^{-3}$ &  6.5 &    60 &    13 & 0.315 & $\pm 0.014^{+0.018}_{-0.020}$\\
15 & 1.6$\cdot 10^{-3}$ &  6.5 &    90 &    14 & 0.262 & $\pm 0.012^{+0.015}_{-0.012}$\\
16 & 2.5$\cdot 10^{-3}$ &  6.5 &   150 &    19 & 0.227 & $\pm 0.009^{+0.011}_{-0.017}$\\
17 & 4.0$\cdot 10^{-3}$ &  6.5 &   250 &    26 & 0.139 & $\pm 0.008^{+0.019}_{-0.007}$\\
18 & 6.3$\cdot 10^{-3}$ & 10 &   450 &    24 & 0.150 & $\pm 0.008^{+0.015}_{-0.008}$\\
19 & 1.0$\cdot 10^{-2}$ & 22 &   800 &    56 & 0.119 & $\pm 0.012^{+0.022}_{-0.018}$\\
20 & 1.6$\cdot 10^{-2}$ &  6.5 &  1200 &    19 & 0.059 & $\pm 0.005^{+0.011}_{-0.022}$\\
21 & 2.5$\cdot 10^{-2}$ & 22 &  1500 &    45 & 0.061 & $\pm 0.008^{+0.020}_{-0.010}$\\
22 & 4.0$\cdot 10^{-2}$ &  6.5 &  2000 &   167 & 0.037 & $\pm 0.007^{+0.019}_{-0.045}$\\
23 & 8.1$\cdot 10^{-2}$ & 10 &  5000 &   156 & 0.018 & $\pm 0.004^{+0.016}_{-0.022}$\\
24 & 0.2 & 90 &  5000 &   388 & 0.008 & $\pm 0.010^{+0.005}_{-0.010}$\\
\hline
\end{tabular}
\end{center}
\caption{Values of the slope $dF_2/d\ln Q^2$ and their errors
calculated from fitting ZEUS $F_2$ data to the form $a+b\ln Q^2$.
The columns labelled $Q^2_{min}, Q^2_{max}$ and $\langle Q^2\rangle$ 
give the minimum, maximum and average values of $Q^2$ in a bin. 
$\langle Q^2\rangle$ is calculated as 
described in \Se{sec:df2dln2} of the text. 
\label{tab:df}
}
\end{table}

\begin{table}[ht]
\begin{center}
\vspace{1.5cm} 
\begin{tabular}{|c|r|r|r|}
\hline
Parameter & $xg(x)$ & $xS(x)$ & $x\Delta_{ud}(x)$ \\
\hline
 $A$           &  1.77 &  0.520 &  6.07  \\
 $\delta $     & -0.225 & -0.241 &  1.27  \\
 $\eta$        &  9.07 &  8.60 &  3.68  \\
 $\varepsilon$ &          &  0.290 &           \\
 $\gamma$      &  3.00 &  8.27 &           \\
\hline
\end{tabular}
\end{center}
\caption{Values for the parameters of the nominal ZEUS NLO QCD fit 
(ZEUSQCD) at the starting scale $Q_0^2=7\unit{GeV}^2$. The parameters 
are defined in \Gl{mbb:param} of the text.
\label{tab:qcd}
}
\end{table}

\newpage
\clearpage
\pagebreak


\newpage

\begin{figure}[ht]
\begin{center}
\epsfig{file=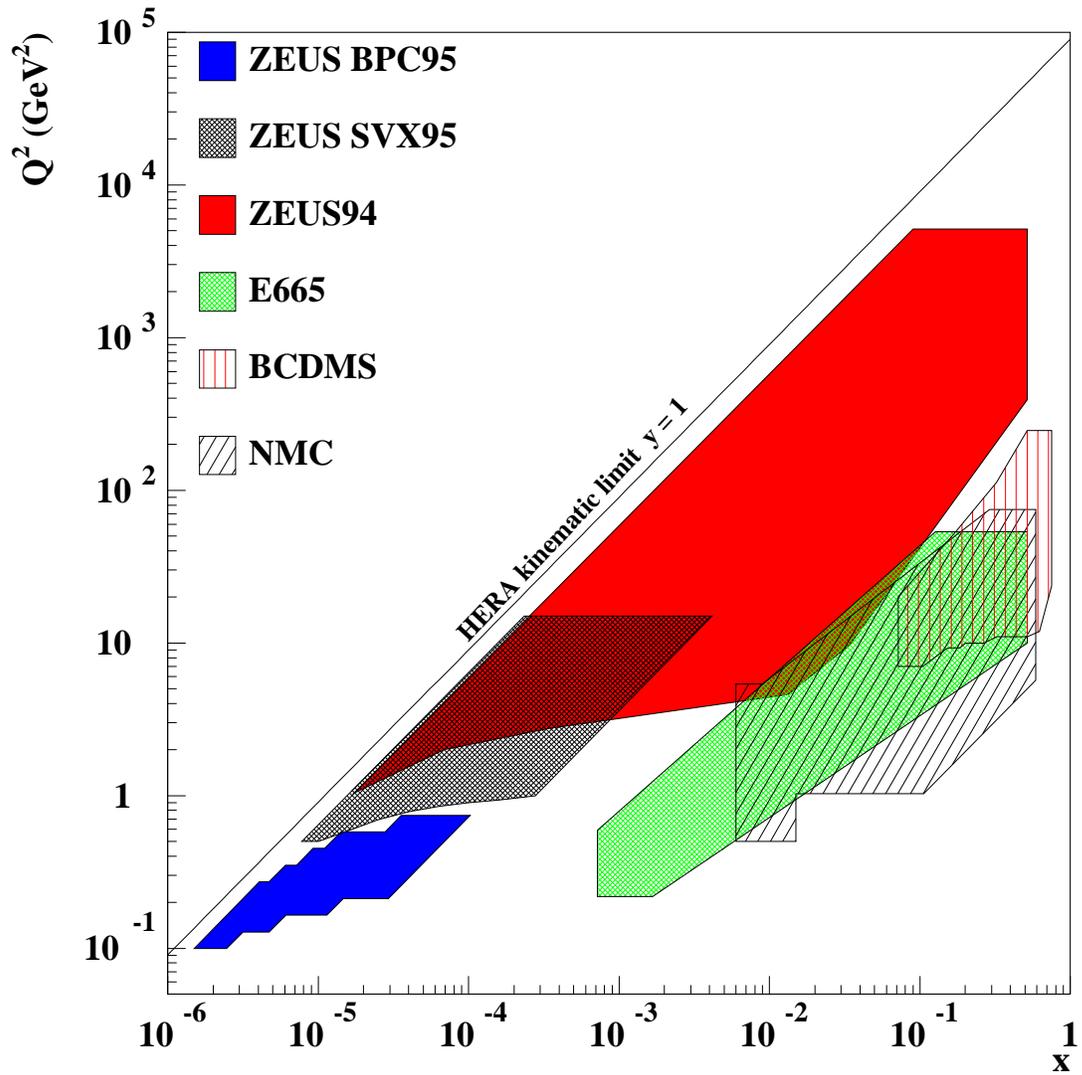,height=16cm}
\end{center}
\caption{The $(x,Q^2)$ plane showing the regions covered by the
ZEUS data sets BPC95, ZEUS94 and SVX95 together with regions
covered by the fixed target experiments E665, BCDMS and NMC. 
}
\label{fig:kinreg}
\end{figure}

\begin{figure}[ht]
\begin{center}
\epsfig{file=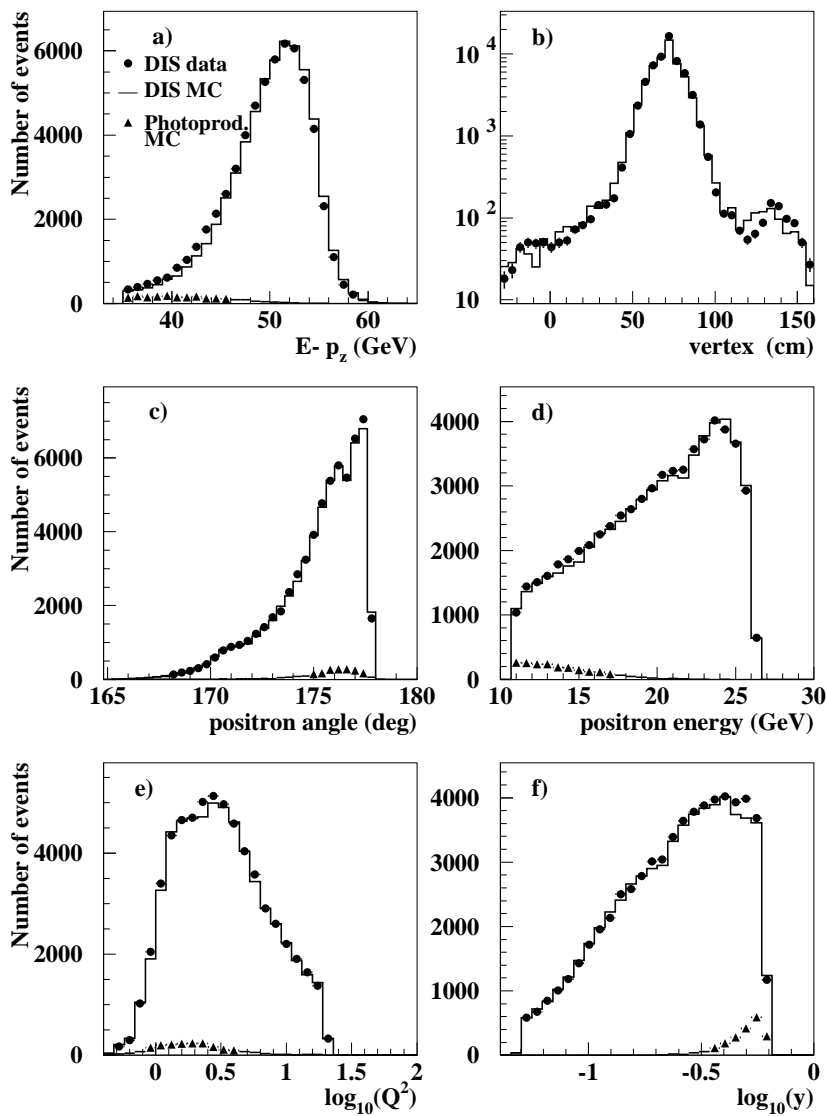,height=16cm}
\end{center}
\caption{Distributions from the SVX95 analysis showing the comparison
between data and simulation: (a) $\delta$ or $E-P_Z$ as defined in 
\Se{sec:svxcuts}; (b) the $Z$-position of the primary vertex; 
(c) the positron scattering
angle $\theta_e$; (d) the energy, $E^\prime_e$, of the scattered positron;
(e) $\log_{10}Q^2$; (f) $\log_{10}y$. In all cases the data are represented
by filled circles, the simulation by the open histograms and the 
photoproduction
background (calculated from the MC described in \Se{sec:svxmc}) by the
filled triangles. Both MC calculations are normalised to the
luminosity of the data.
}
\label{fig:svxplots}
\end{figure}

\begin{figure}[ht]
\begin{center}
\epsfig{file=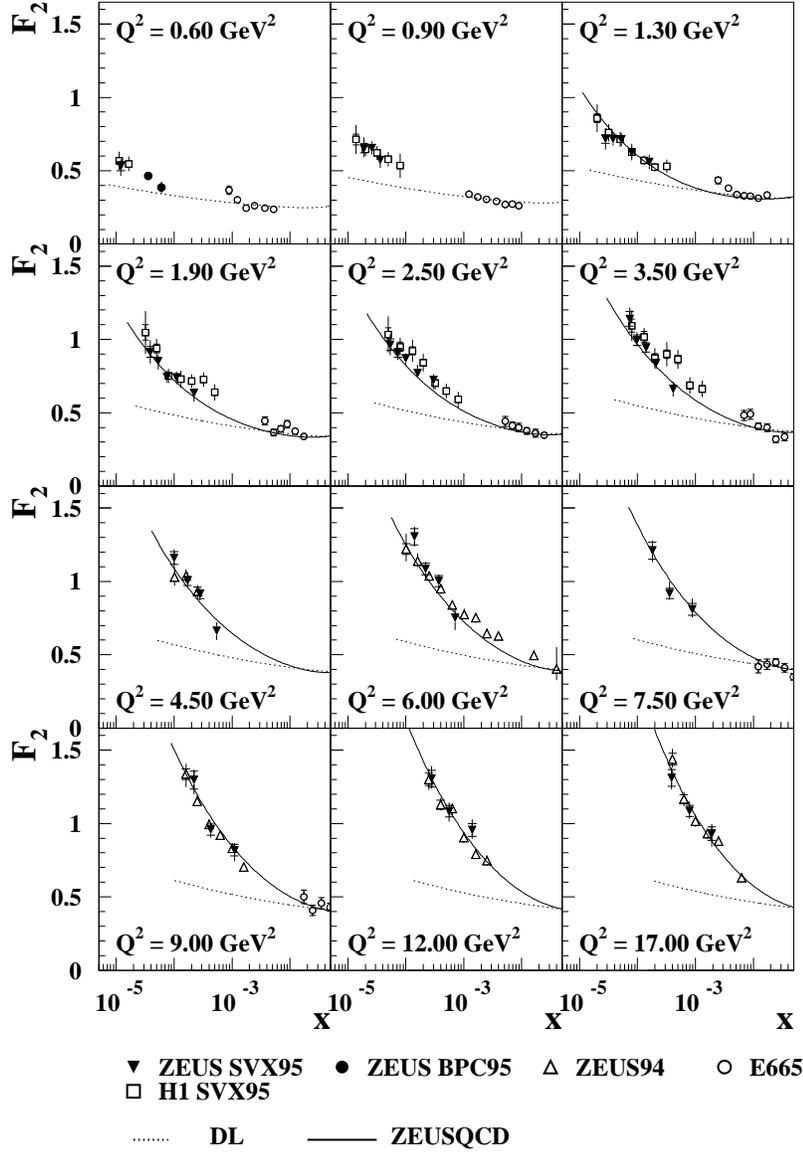,height=16cm}
\end{center}
\caption{
The ZEUS SVX95 $F_2$ data as a function of $x$ for different $Q^2$ bins
together with previous ZEUS data (ZEUS 1994 \cite{ZF2}, 
ZEUS BPC 1995 \cite{ZBPC}), data from H1 SVX95 \cite{H1_svx95} and fixed 
target data (E665 \cite{E665F2}). Error bars 
correspond to the statistical and systematic errors added in quadrature.  
The overall normalisation errors are not shown. The curves shown are 
(dotted) the Donnachie-Landshoff Regge model \cite{dltwo} and 
(full) the ZEUS NLO QCD fit.}
\label{fig:svtx1}
\end{figure}

\begin{figure}[ht]
\begin{center}
\epsfig{file=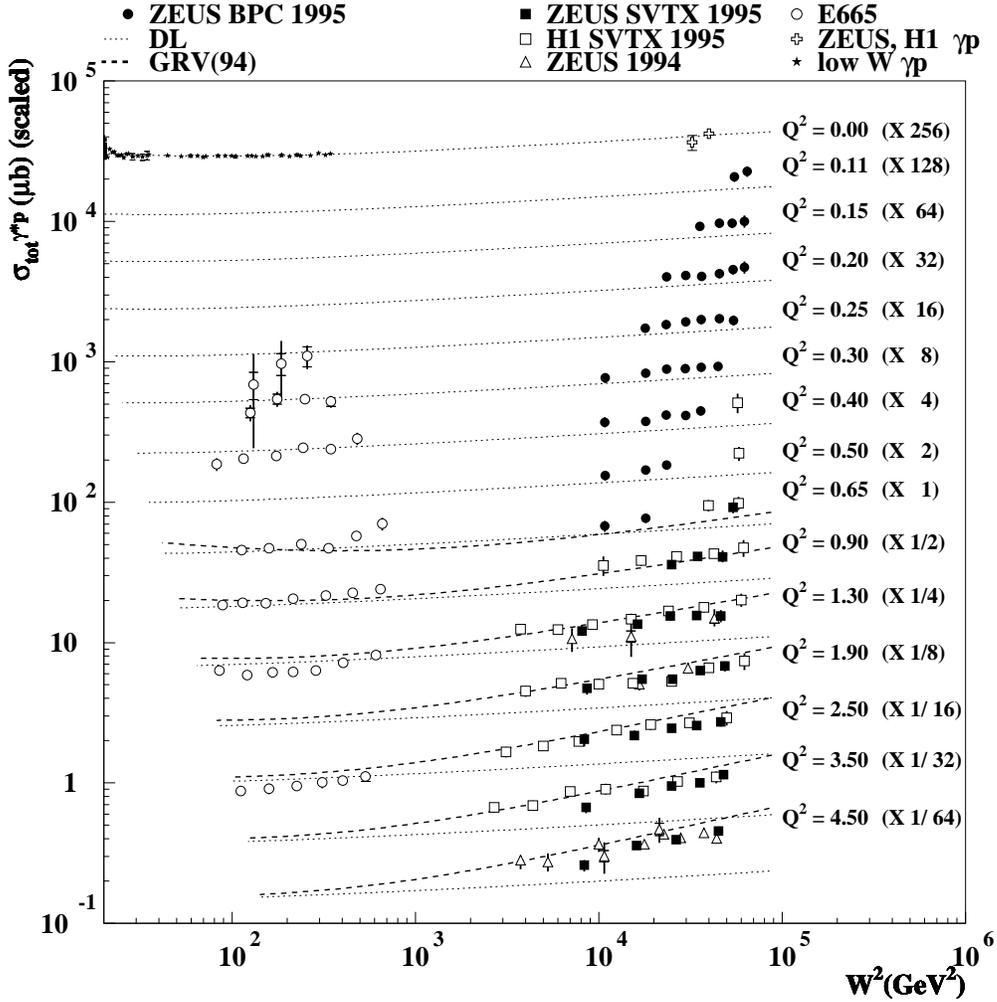,height=15cm}
\end{center}
\caption{
The total $\gamma^*p$ cross-section, $\sigma_{\rm tot}^{\gamma^*p}$, 
as a function of $W^2$ at different $Q^2$ [GeV$^2$]. The data of this analysis 
(ZEUS SVX95) are shown together with previous ZEUS and H1 data 
(ZEUS94, ZEUS BPC95, H1 SVX95). Also the total photoproduction 
cross-sections from ZEUS and H1 and from fixed target experiments at 
lower $W$ are shown. 
Predictions from the DL \cite{dltwo} and GRV94 \cite{grv94} models are 
indicated by the dotted and dashed curves, respectively.
}
\label{fig:svtx2}
\end{figure}

\begin{figure}[ht]
\begin{center}
\epsfig{file=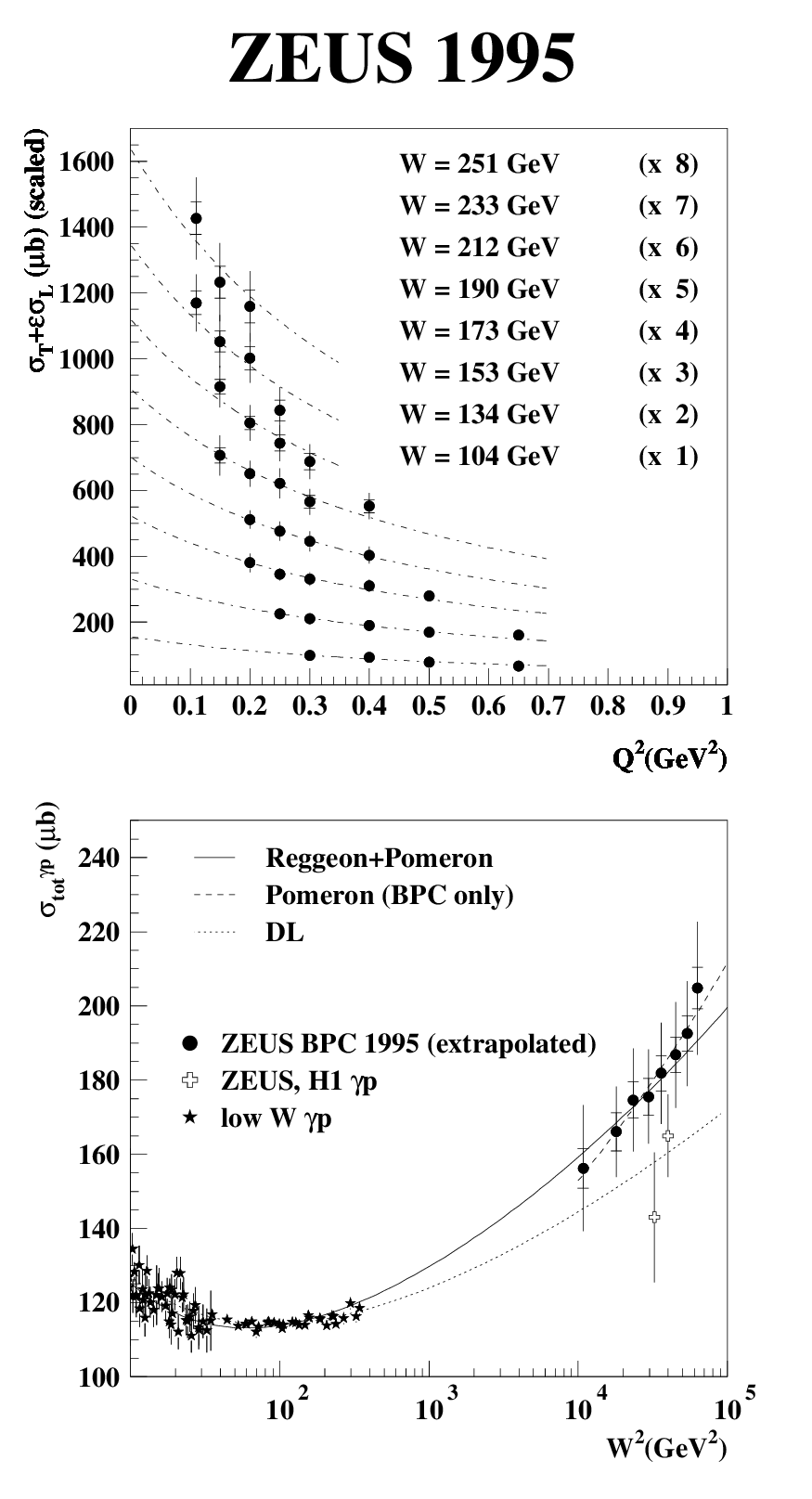,height=16cm}
\end{center}
\caption{
{\it Upper plot.} ZEUS BPC measurements of the total cross-section 
$\sigma_T + \epsilon \sigma_L$ in bins of $W$ as a function of $Q^2$ and 
the GVMD fit to the data.
{\it Lower plot.} $\sigma_{\rm tot}^{\gamma p}$ as a function of $W^2$.
The ZEUS BPC95 points are those from the GVMD extrapolation
($\sigma_{0}^{\gamma p}$). Also
shown are direct measurements of the total photoproduction
cross-section from H1, ZEUS and earlier experiments at low energies.
The curves show Regge fits: the original DL fit \cite{dltwo}
to the low $W$ data (dotted); the Pomeron only fit to the 
BPC $\sigma_{0}^{\gamma p}$ data (dashed) and the Pomeron+Reggeon
fit to the low $W$ and BPC $\sigma_{0}^{\gamma p}$ data (full).
}
\label{fig:bpcvmd}
\end{figure}

\begin{figure}
\begin{center}
\epsfig{file=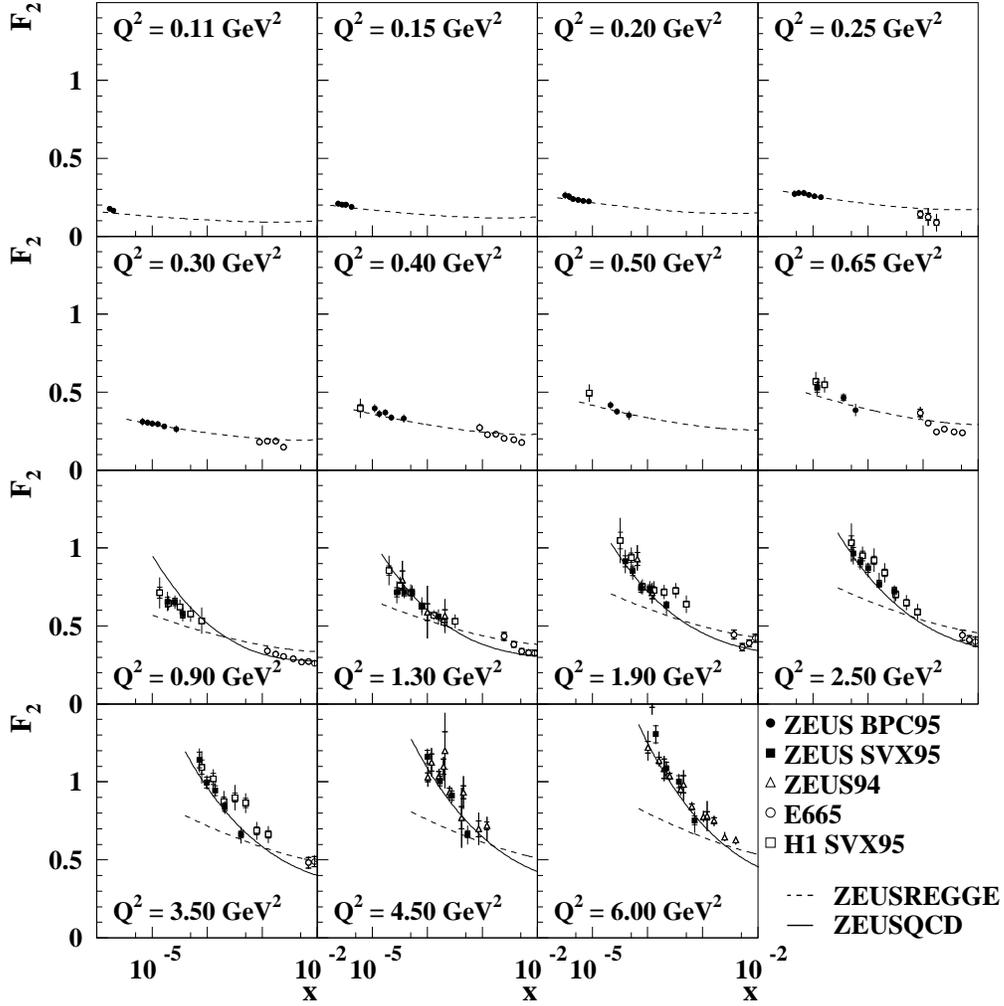,height=15cm}
\end{center}
\caption{
Low $Q^2$ $F_2$ data for different $Q^2$ bins together with the ZEUS
Regge fit (dashed curves) to the BPC95 data as described in 
\Se{sec:lowq2reg}. Also shown at larger values of $Q^2$ is the ZEUS NLO 
QCD fit (full curves) as described in \Se{sec:qcd}.
}
\label{fig:dlfit}
\end{figure}

\begin{figure}[ht]
\begin{center}
\epsfig{file=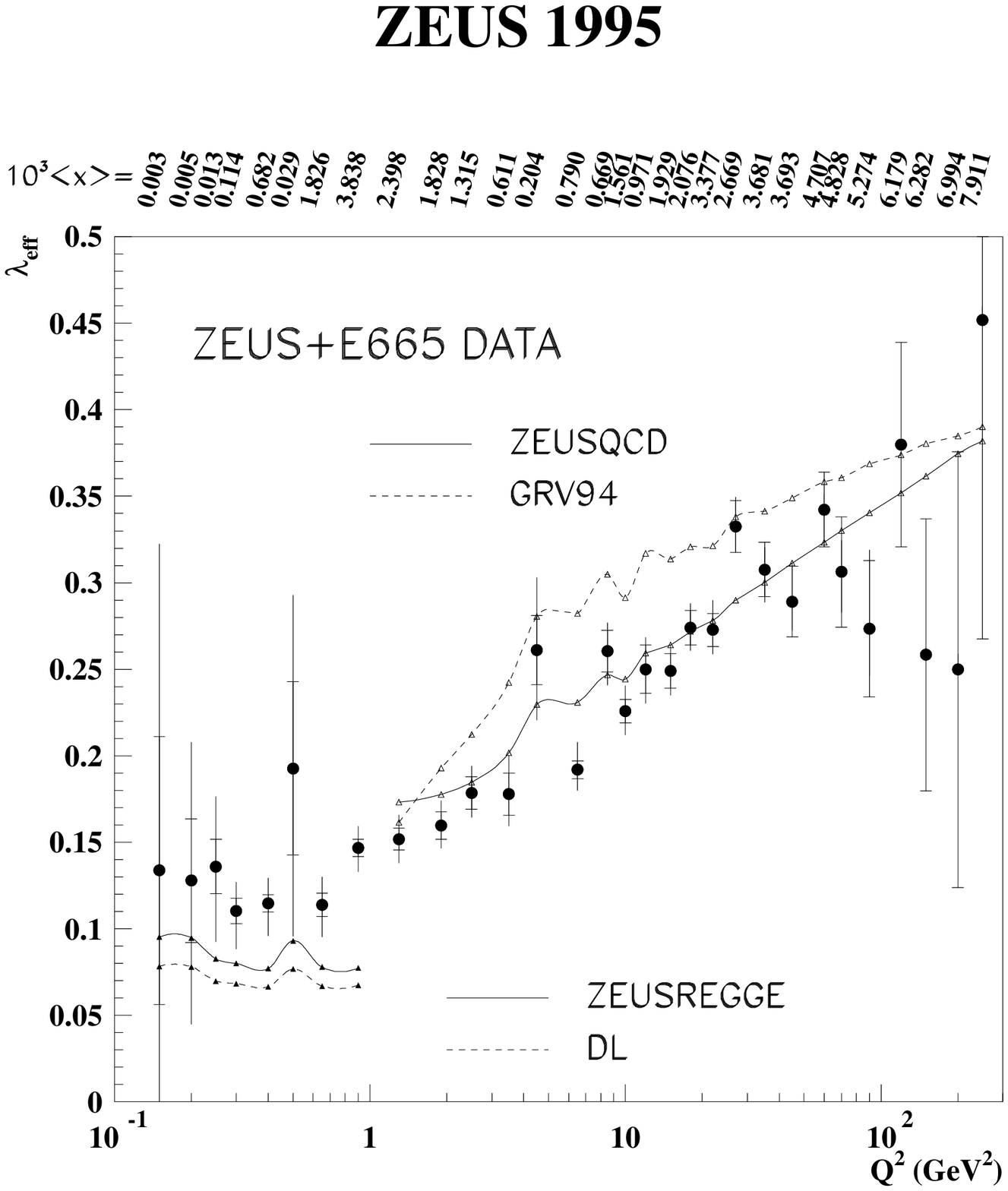,width=14cm}
\end{center}
\caption{$\lambda_{eff}=d\ln F_2/d\ln(1/x)$ as a function of $Q^2$ calculated
by fitting $F_2=A x^{-\lambda_{eff}}$ to ZEUS and E665 data with $x<0.01$. 
The inner error bar shows the statistical error and the outer the total 
statistical and systematic error added in quadrature.  
$\langle x \rangle$ is calculated as described in \Se{sec:lameff}. 
The DL and GRV94 calculations, shown as points linked by dashed lines,  
are from the Donnachie-Landshoff Regge fit \cite{dltwo} and the GRV94 
NLO QCD fit \cite{grv94}, respectively. The ZEUSREGGE and ZEUSQCD 
calculations, shown as points linked by full lines,
are from the ZEUS Regge and NLO QCD fits described in Secs 
\ref{sec:lowq2reg} and \ref{sec:qcd} of the text respectively.
In all cases the points are obtained using the same
weighted range of $x$ as for the experimental data.
}
\label{fig:lambda_eff}
\end{figure}

\begin{figure}[ht]
\begin{center}
\epsfig{file=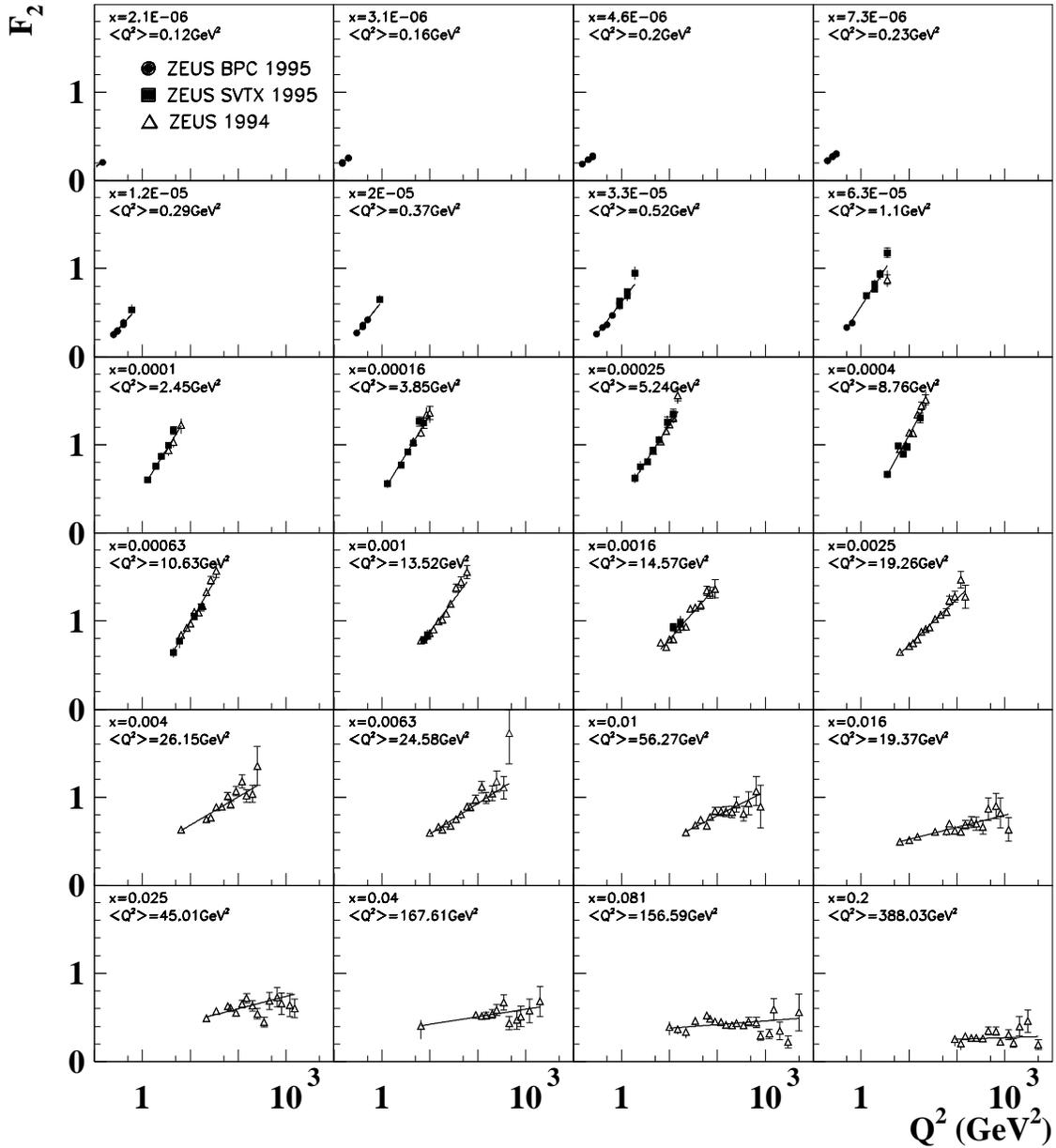,height=18cm}
\end{center}
\caption{$F_2$ as a function of $Q^2$ in bins of $x$ from the
ZEUS data sets BPC95, SVX95, ZEUS94. The linear fits $F_2=a+b\ln Q^2$
are also shown. The values of $\langle Q^2\rangle$ 
given in the plots are calculated as described in \Se{sec:df2dln2}.
}
\label{fig:f2vq_binx}
\end{figure}

\begin{figure} [ht]
\begin{center}
\epsfig{file=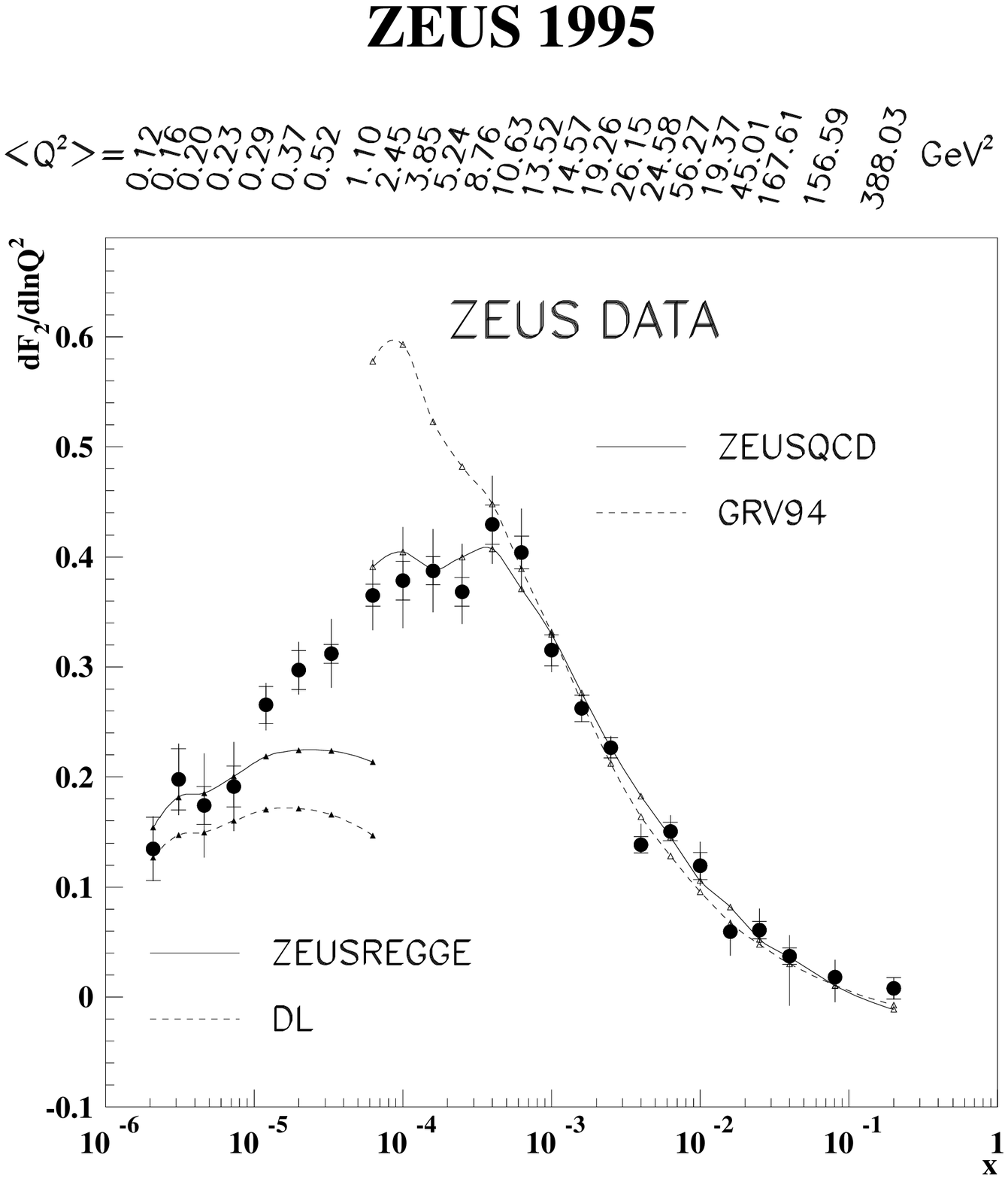,width=14cm}
\end{center}
\caption{$d F_2/d\ln Q^2$ as a function of $x$ calculated by fitting
ZEUS $F_2$ data in bins of $x$ to the form $a+b\ln Q^2$. The inner error bar
shows the statistical error and the outer the total statistical
and systematic error added in quadrature.  $\langle Q^2 \rangle$ is 
calculated as described in \Se{sec:df2dln2}. 
The DL and GRV94 calculations, shown as points linked by dashed lines,  
are from the Donnachie-Landshoff Regge fit \cite{dltwo} and the GRV94 
NLO QCD fit \cite{grv94}, respectively. The ZEUSREGGE and ZEUSQCD 
calculations, shown as points linked by full lines,
are from the ZEUS Regge and NLO QCD fits described in Secs 
\ref{sec:lowq2reg} and \ref{sec:qcd} of the text respectively.
In all cases the points are obtained using the same
weighted range of $Q^2$ as for the experimental data.
}
\protect\label{fig:df2dlnq2}
\end{figure}

\begin{figure}[ht]
\begin{center}
\epsfig{file=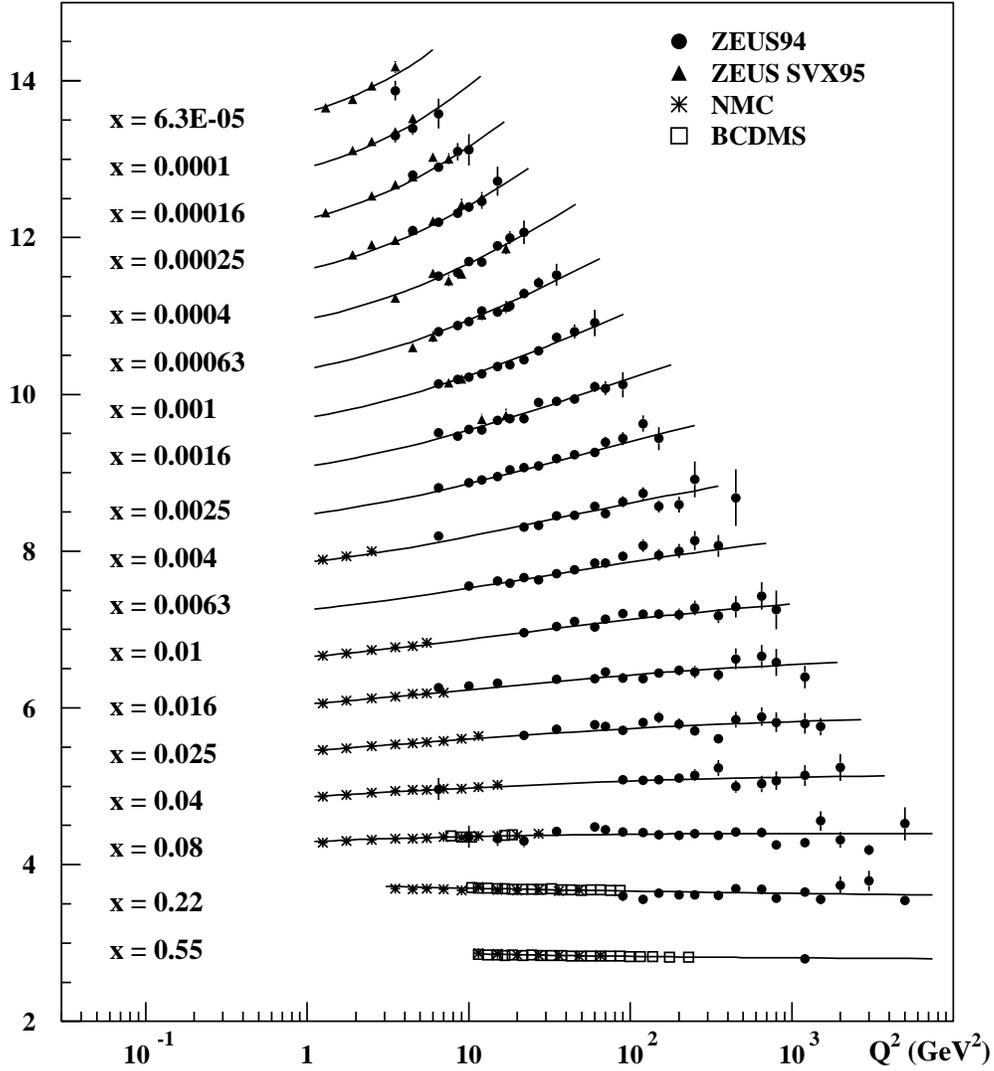,height=15cm}
\end{center}
\caption{
The proton structure function $F_2$ versus $Q^2$
at fixed values of $x$. Data are from the ZEUS94 and SVX95 analyses
and from the NMC and BCDMS fixed target experiments.
The solid lines correspond to the QCD fit described in the text. For 
clarity an amount $C_i = 13.6 - 0.6 i$ is added to $F_2$ where 
$i = 1$ (18) for the lowest (highest) $x$ value.}
\protect\label{fig:f2q2fit}
\end{figure}

\begin{figure} [ht]
\begin{center}
\epsfig{file=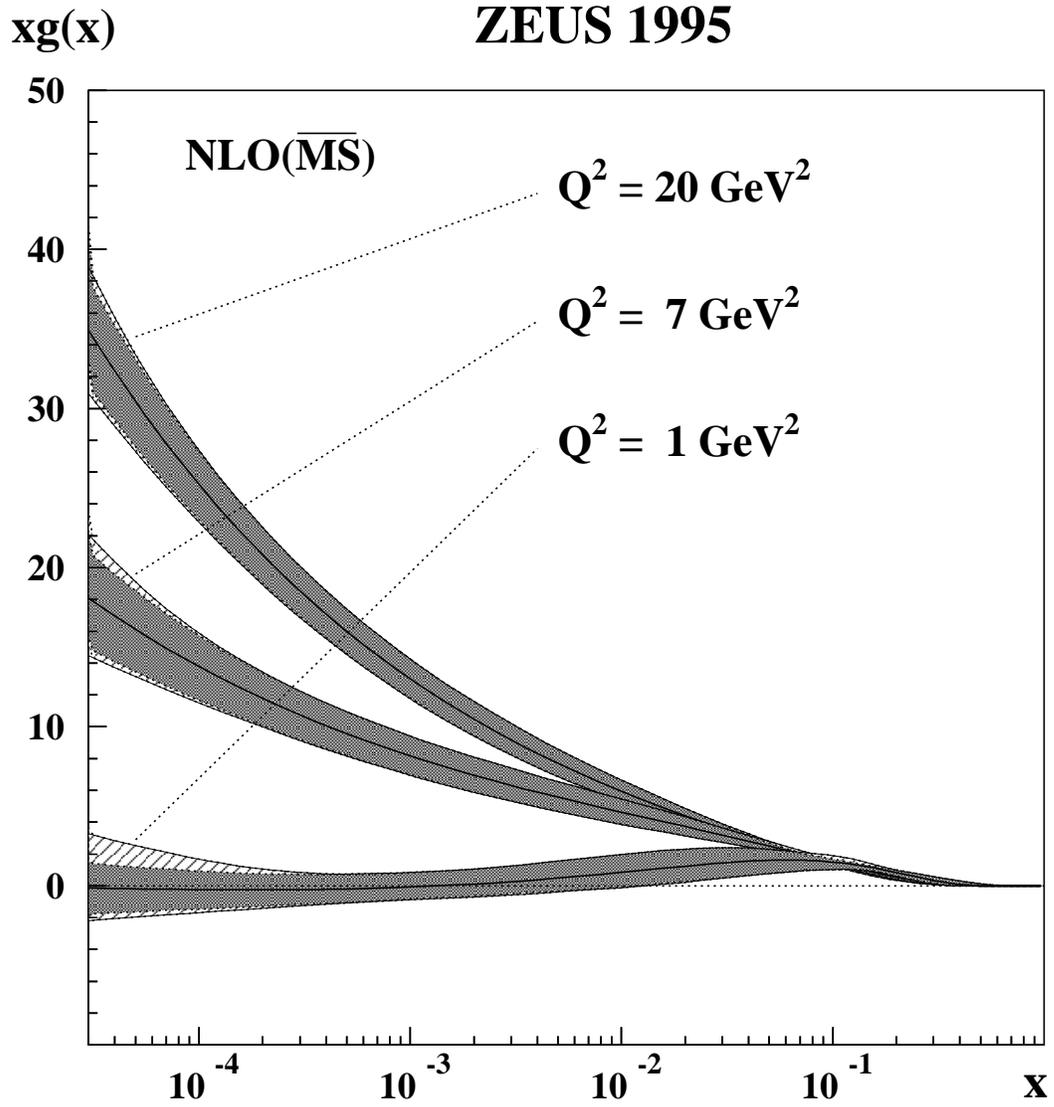,width=14cm}
\end{center}
\caption{
The gluon momentum distribution $xg(x)$ as a function of $x$ at fixed
values of $Q^2 = 1$, 7 and 20~\gev ~from the ZEUS QCD fit. 
The inner shaded bands show the
`HERA standard' errors of \Se{sec:qcd}. The outer hatched
bands indicate the quadratic sum of the `HERA standard' and
the `parameterisation' errors.
}
\protect\label{fig:xg1720}
\end{figure}

\begin{figure} [ht]
\begin{center}
\epsfig{file=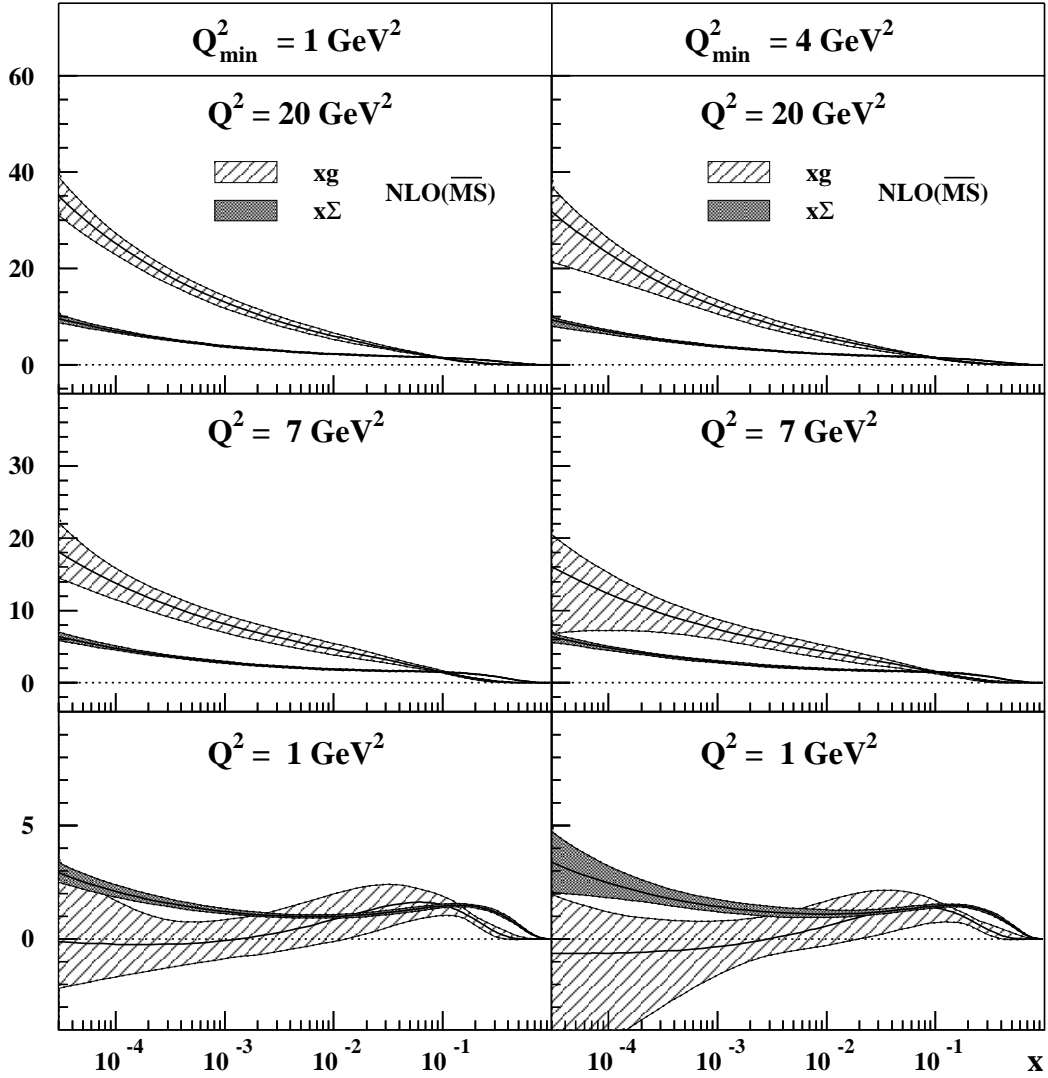,width=14cm}
\end{center}
\caption{The quark singlet momentum distribution, $x\Sigma$ (shaded), and 
the gluon momentum distribution, $xg(x)$ (hatched), as functions of $x$ 
at fixed values of $Q^2 = 1$, 7 and 20~\gev ~from the ZEUS QCD fit. 
The error bands correspond to the
quadratic sum of all error sources considered for each parton density.
The three left-hand plots show the results from the standard fit
of \Se{sec:qcd} including $F_2$ data with $Q^2 > 1\unit{GeV}^2$; the 
three right-hand plots show the corresponding results
from the fit described in \Se{sec:loq2qcd} for which the data must satisfy
$Q^2 > 4\unit{GeV}^2$.
}
\protect\label{fig:xSxg}
\end{figure}

\end{document}